\documentclass[epj, nopacs]{svjour}
\usepackage[dvips]{graphicx}
\usepackage{amsfonts}
\usepackage{amssymb}
\usepackage{epsfig}
\usepackage{cite}
\usepackage{subfigure}
\usepackage{rotating}
\DeclareGraphicsExtensions{.eps}

\newcommand{\newc}{\newcommand}

\newc{\MZ}{M_Z}
\newc{\bbbar}{b\bar{b}}
\newc{\ccbar}{c\bar{c}}
\newc{\ssbar}{s\bar{s}}
\newc{\ttbar}{t\bar{t}}

\newc{\lag}{\mathcal{L}}
\newc{\mM}{\mathcal{M}}
\newc{\lam}{\lambda}
\newc{\intd}{\int\frac{d^Dq}{(2\pi)^D}}

\newc{\MET}{{\not\!\!E_T}}
\newc{\pt}{p_{T}}
\newc{\ptlep}{p^l_T}
\newc{\ptb}{p^b_T}
\newc{\Dphi}{\Delta\phi}

\newc{\pdfannotlink}{\pdfstartlink}
\newc{\be}{\begin{equation}}
\newc{\ee}{\end{equation}}
\newc{\bea}{\begin{eqnarray}}
\newc{\eea}{\end{eqnarray}}
\newc{\eqa}{\begin{eqnarray}}
\newc{\qea}{\end{eqnarray}}
\newc{\mc}{\mathcal}
\newc{\gsim}{\gtrsim}
\newc{\lsim}{\lesssim}
\newc{\mb}{\textrm{ mb}}
\newc{\Wp}{W^+}
\newc{\Wm}{W^-}
\newc{\Zgam}{Z(\gamma^*)}
\newc{\Wpm}{W^{\pm}}
\newc{\numu}{\nu_{\mu}}
\newc{\order}{\mathcal{O}}
\newc{\sig}{\sigma}
\newc{\sigh}{\hat{\sigma}}
\newc{\sigeff}{\sigma_{\textrm{eff}}}
\newc{\sigeffN}{\sigma_{N,\textrm{eff}}}
\newc{\as}{\alpha_S}
\newc{\aw}{\alpha}
\newc{\roots}{\sqrt{s}}
\newc{\bmix}{B^0\textrm{-}\bar{B}^0}
\newc{\barn}{\textrm{b}}
\newc{\BR}{\mathcal{BR}}
\newc{\nn}{\langle n\rangle}

\newc{\herwig}{\texttt{HERWIG6.510}}
\newc{\mcfm}{\texttt{MCFM}}
\newc{\mstwlo}{\texttt{MSTW08 LO}}
\newc{\vegas}{\texttt{VEGAS}}
\newc{\madgraph}{\texttt{MADGRAPH}}
\newc{\legacy}{\texttt{LEGACY}}

\begin{document}

\title{Same-sign W pair production as a probe of double parton
  scattering at the LHC}

\author{Jonathan~R.~Gaunt\inst{1}\and Chun-Hay~Kom\inst{1}\and
  Anna~Kulesza\inst{2}\and W.~James~Stirling\inst{1}}

\institute{Cavendish Laboratory, J.J. Thomson Avenue, Cambridge CB3
  0HE, UK \and Institute for Theoretical Particle Physics and
  Cosmology, RWTH Aachen University D-52056 Aachen, Germany}

\date{\today}

\abstract{ We study the production of same-sign $W$ boson pairs at the
  LHC in double parton interactions.  Compared with simple factorised
  double parton distributions (dPDFs), we show that the recently
  developed dPDFs, GS09, lead to non-trivial kinematic correlations
  between the $W$ bosons.  A numerical study of the prospects for
  observing this process using same-sign dilepton signatures,
  including $\Wpm\Wpm jj$, di-boson and heavy flavour backgrounds, at
  14~TeV centre-of-mass energy is then performed.  It is shown that a
  small excess of same-sign dilepton events from double parton
  scattering over a background dominated by single scattering
  $\Wpm\Zgam$ production could be observed at the LHC.  
}


\maketitle


\section{Introduction}

The Large Hadron Collider (LHC) at CERN will offer many interesting
tests of Standard Model (SM) physics and, indeed, a precise knowledge
of SM processes is necessary to identify unambiguously New Physics
(NP) signals from SM backgrounds. The high LHC collision energy will
open up a new kinematic regime in which certain SM processes will
become precisely measureable for the first time. An example is
multiple parton hard-scattering, i.e.  events in which two or more
distinct hard parton interactions occur simultaneously.  The
theoretical study of such processes goes back to the early days of the
parton model \cite{Landshoff:1978fq, Takagi:1979wn ,Goebel:1979mi}
with subsequent extension to perturbative QCD \cite{Paver:1982yp,
  Humpert:1983pw, Mekhfi:1983az, Humpert:1984ay, Ametller:1985tp,
  Mekhfi:1985dv, Halzen:1986ue, Sjostrand:1987su, Mangano:1988sq,
  Godbole:1989ti, Drees:1996rw, Calucci:1997uw,
  Calucci:1999yz}. Experimental evidence for double parton scattering
(DPS) has been found in $\sqrt{s} = 63$~GeV $pp$ collisions by the AFS
collaboration at the CERN ISR \cite{Akesson:1986iv}, and more recently
in $\sqrt{s} = 1.8$~TeV $p\bar{p}$ collisions by the CDF collaboration
\cite{Abe:1997xk} and $\sqrt{s} = 1.96$~TeV $p\bar{p}$ collisions by
the D0 collaboration \cite{Abazov:2009gc} at the Fermilab Tevatron.

In the standard framework for calculating inclusive hard-scattering
cross sections in hadron-hadron collisions, it is assumed that only
one hard interaction occurs per collision (plus multiple soft
interactions). This assumption is typically justified on the grounds
that the probability of a hard parton-parton interaction in a
collision is very small. Thus the probability of having two or more
hard interactions in a collision is highly suppressed with respect to
the single interaction probability.

However, as the collider centre-of-mass energy becomes larger, we may
expect multiple hard parton collisions to become more important. This
can be understood as follows. Consider a final state consisting of the
products of two hard collisions $A$ and $B$, where for example $A,B =
W, Z, jj, t\bar t, ...$ etc.  It is commonly assumed that double
parton scattering cross sections $\sigma^{DPS}_{(A,B)}$ can be
approximately factorised into the product of two single scattering
cross sections:
 \begin{equation}
\sigma^{DPS}_{(A,B)} =\frac{m}{2}
\frac{\sigma^S_{(A)}\sigma^S_{(B)}}{\sigma_{\rm eff}}
\label{eq:naivefact}
\end{equation}
where the quantity $m$ is a symmetry factor that equals $1$ if $A=B$
and $2$ otherwise.  The factor $\sigeff$ in the denominator has the
dimensions of a cross section.  The reason for this is that given that
one hard scattering occurs, the probability of the other hard
scattering is proportional to the flux of accompanying partons; these
are confined to the colliding protons, and therefore their flux should
be inversely proportional to the area (cross section) of a proton.  Of
course the $(A,B)$ final state can also be produced by a single parton
scattering, with cross section $\sigma^S_{(AB)}$. If the masses of the
final states $A$ and $B$ are fixed, then the collider energy
($\sqrt{s}$) dependence of the hadronic cross sections is controlled
by the $x$ dependence of the parton distribution functions
(PDFs). Because the PDFs increase rapidly with $x$ as $x \to 0$, the
cross sections increase with $\sqrt{s}$.  This implies that
$\sigma^D_{(A,B)}$, which is proportional to a product of single
scattering cross sections, will increase more rapidly with $\sqrt{s}$
than $\sigma^S_{(AB)}$. This raises the possibility that multiple hard
parton scattering, which has received relatively little attention to
date, could provide important backgrounds to NP signals at the LHC
\cite{DelFabbro:1999tf, DelFabbro:2002pw, Hussein:2006xr,
  Hussein:2007gj}. To take a simple example for SM Higgs production,
the `standard' irreducible background to associated $Z+ H(\to b \bar
b)$ production is $q\bar q, gg \to Z b \bar b$. However, there is an
additional DPS background with $ A=Z$ and $B = b \bar b$.

Of course the magnitude of the DPS cross section is directly dependent
on the size of $ \sigma_{\rm eff}$ in Eq.~(\ref{eq:naivefact}).  The
recent CDF and D0 measurements \cite{Abe:1997xk,Abazov:2009gc} at the
Tevatron, utilising the $\gamma + 3$jet final state with $A$
corresponding to $\gamma +$jet production and $B$ to dijet production,
suggest $\sigma_{\rm eff} \sim 15$~mb, which is roughly $20\%$ of the
total (elastic $+$ inelastic) $p\bar{p}$ cross section at the Tevatron
collider energy. The non-perturbative physics that determines $
\sigma_{\rm eff}$ is not well enough understood to be able to predict
its value at the LHC. If it is proportional to the total inelastic
cross section, one might expect a slightly higher value at LHC
energies. Given this uncertainty, the best approach is clearly to find
a {\em benchmark} double scattering process for the LHC, from which $
\sigma_{\rm eff}$ can be determined. This will not only serve to
calibrate DPS backgrounds to NP processes, but will also provide
important information on the non-perturbative structure of the proton.

In fact, Eq.~(\ref{eq:naivefact}) is only an approximation to the
following more general expression:
\begin{eqnarray}
\sigma^{DPS}_{(A,B)} &=& \frac{m}{2}\sum_{i,j,k,l}\int
dx_1dx_2dx_1'dx_2'd^2b \nonumber \\ &&
\times \Gamma_{ij}(x_1,x_2,b;t_1,t_2)\; \Gamma_{kl}(x_1',x_2',b;t_1,t_2)
\nonumber \\ && \times
\hat{\sigma}^A_{ik}(x_1,x_1')\hat{\sigma}^B_{jl}(x_2,x_2') .
\end{eqnarray}
The $\Gamma_{ij}(x_1,x_2,b;t_1,t_2)$ represent generalised double
parton distributions. They may be loosely interpreted as the inclusive
probability distributions to find a parton $i$ with longitudinal
momentum fraction $x_1$ at scale $t_1\equiv \ln(Q_1^2)$ in the proton,
in addition to a parton $j$ with longitudinal momentum fraction $x_2$
at scale $t_2\equiv \ln(Q_2^2)$, with the two partons separated by a
transverse distance $b$. The scale $t_1$ is given by the
characteristic scale of subprocess $A$, whilst $t_2$ is equal to the
characteristic scale of subprocess $B$.

It is typically taken that $\Gamma_{ij}(x_1,x_2,b;t_1,t_2)$ may be
decomposed in terms of longitudinal and transverse components as
follows:
\begin{equation}
\Gamma_{ij}(x_1,x_2,b;t_1,t_2) = D^{ij}_h(x_1,x_2;t_1,t_2)F^i_j(b).
\end{equation}

Making the further assumption that $F^i_j(b)$ is the same for all
parton pairs $ij$ involved in the DPS of interest, this leads to:
\begin{eqnarray}
\sigma^{DPS}_{(A,B)} &=& \frac{m}{2\sigma_{\rm eff}} \sum_{i,j,k,l}
\int dx_1dx_2dx_1'dx_2' \nonumber \\ & & \times
D^{ij}_p(x_1,x_2;t_1,t_2)\; D^{kl}_p(x_1',x_2';t_1,t_2) \nonumber \\ &
& \times \hat{\sigma}^A_{ik}(x_1,x_1')\hat{\sigma}^B_{jl}(x_2,x_2') ,
\nonumber \\ \sigma_{\rm eff} &=& \left[\int d^2b (F(b))^2\right]^{-1}
.
\label{eq:DPSmaster}
\end{eqnarray}

If one ignores longitudinal momentum correlations such that the
components $D^{ij}_h$ take the form $D^{ij}_h(x_1,x_2;t_1,t_2)$ $=$
$D^{i}_h(x_1;t_1)D^{j}_h(x_2;t_2)$, then one finally arrives at the
form of Eq.~(\ref{eq:naivefact}). This is the approach that has been
taken in existing phenomenological calculations regarding DPS
\cite{DelFabbro:1999tf, DelFabbro:2002pw, Hussein:2006xr,
  Hussein:2007gj, Kulesza:1999zh, Maina:2009vx, Maina:2009sj,
  Berger:2009cm}.  Such an approximation is typically justified at low
$x$ values on the grounds that the population of partons is large at
these values.

On the other hand, a number of theoretical studies
\cite{Snigirev:2003cq, Korotkikh:2004bz,Cattaruzza:2005nu,
  Gaunt:2009re} have suggested that non-negligible longitudinal
momentum correlations do exist in the double parton distributions
(dPDFs) $D^{ij}_h(x_1,x_2;t_1,t_2)$.  These papers have investigated
the special case in which the two factorisation scale arguments of the
dPDFs are set equal $t_1 = t_2 = t$. In \cite{Snigirev:2003cq,
  Korotkikh:2004bz, Cattaruzza:2005nu}, it is shown that pQCD (`double
DGLAP') evolution causes the dPDFs to deviate from factorised forms,
such that even if factorised forms are a good approximation at low
scales, they cannot be so at higher scales. Ref.~\cite{Gaunt:2009re}
goes further, and shows that as a result of the sum rules the dPDFs
have to obey, the PDF factorisation hypothesis $D^{ij} = D^{i}D^{j}$
cannot hold for any $i,j$ at any scale $t$ -- although it may be a
reasonable approximation for sea quark and gluon distributions at
small $x$.

\begin{figure*}[!ht]
  \begin{center}
    \scalebox{1.2}{
      \includegraphics{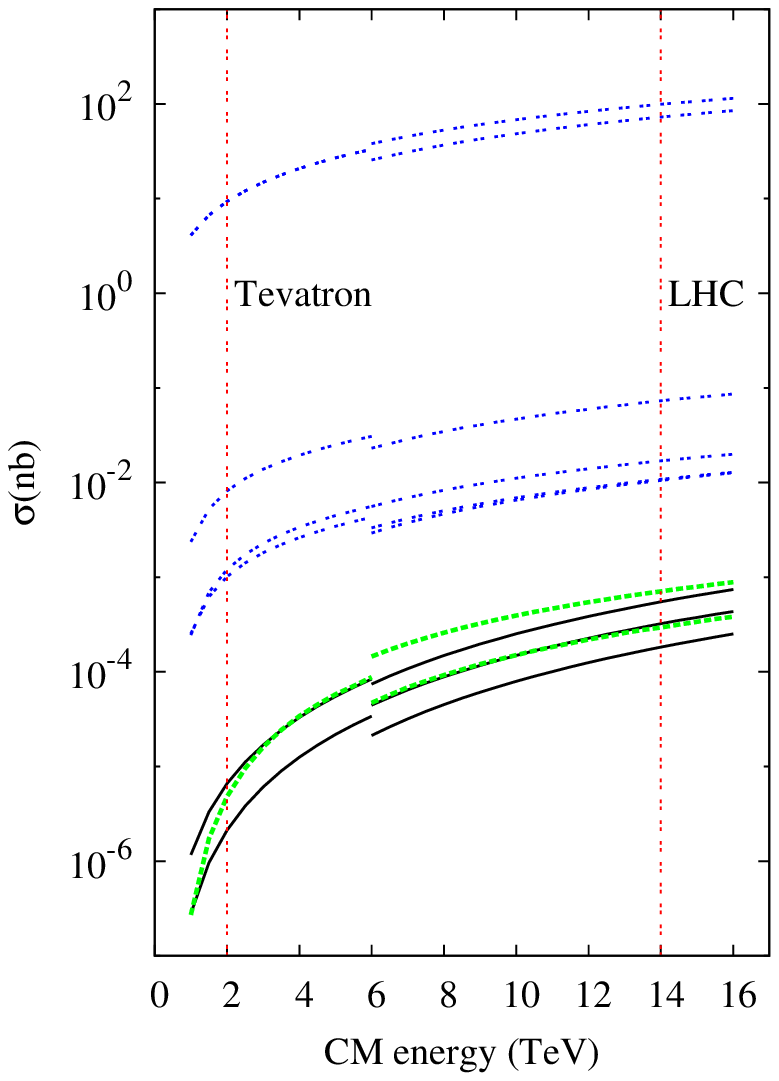}
    }
    \put(-120,335){\tiny $\sigma_{\Wp}$}
    \put(-120,315){\tiny $\sigma_{\Wm}$}
    \put(-120,235){\tiny $\sigma_{\Wp\Wm}$}
    \put(-120,215){\tiny $\sigma_{\Wp Z}$}
    \put(-120,190){\tiny $\sigma_{ZZ}$, $\sigma_{\Wm Z}$}
    \put(-120,162){\rotatebox{12}{\tiny $\sigma_{\Wp\Wp jj}$}}
    \put(-120,147){\rotatebox{15}{\tiny $\sigma_{\Wm\Wm jj}$}}
    \put(-20,170){\tiny{$\sigma^{DPS}_{\Wp\Wm}$}}
    \put(-20,160){\tiny{$\sigma^{DPS}_{\Wp\Wp}$}}
    \put(-20,150){\tiny{$\sigma^{DPS}_{\Wm\Wm}$}}
    \caption{Cross sections of various processes in proton
      (anti-)proton collisions as a function of $\sqrt{s}$.  The
      dotted curves correspond to single scattering processes, while
      the solid curves correspond to double scattering processes
      computed using GS09 dPDFs.  From top down, the $p\bar{p}$ single
      scattering cross sections in the Tevatron region are
      $\sigma_{\Wpm}$, $\sigma_{\Wp\Wm}$, $\sigma_{\Wpm Z}$,
      $\sigma_{ZZ}$ and $\sigma_{\Wpm\Wpm jj}$.  The double scattering
      cross sections are $\sigma^{DPS}_{\Wp\Wm}$ and
      $\sigma^{DPS}_{\Wpm\Wpm}$.  In the LHC region, the dotted curves
      are $pp$ cross sections of $\sigma_{\Wp}$, $\sigma_{\Wm}$,
      $\sigma_{\Wp\Wm}$, $\sigma_{\Wp Z}$, $\sigma_{ZZ}$, $\sigma_{\Wm
        Z}$, $\sigma_{\Wp\Wp jj}$ and $\sigma_{\Wm\Wm jj}$, while the
      solid curves are $\sigma^{DPS}_{\Wp\Wm}$,
      $\sigma^{DPS}_{\Wp\Wp}$ and $\sigma^{DPS}_{\Wm\Wm}$.}
    \label{fig:xsecProc} 
  \end{center}
\end{figure*}

Thus, a more accurate way to model DPS is to use
Eq.~(\ref{eq:DPSmaster}) along with a set of dPDFs which incorporates
the effects of pQCD evolution and sum rule constraints (e.g. GS09
\cite{Gaunt:2009re}).  The practical implication of the longitudinal
momentum correlations inherent in the GS09 set is that the final
states $A$ and $B$ will necessarily be correlated in longitudinal
momentum, and in particular the rapidity distribution of $A$ will not
be independent of that of $B$.

In this study we focus on same-sign $W$ pair production at the LHC as
the paradigm DPS benchmark process, i.e. $A = B = W^+$ and $A = B =
W^-$. In this context, $W^\pm W^\pm$ production was first discussed in
Ref.~\cite{Kulesza:1999zh} and subsequently studied in more detail in
Ref.~\cite{Cattaruzza:2005nu} and Refs.~\cite{Maina:2009vx,
  Maina:2009sj}, the latter of which also compared the cross sections
of DPS $\Wpm\Wpm$ and $\Wpm\Wpm jj$ in single scattering with the
inclusion of kinematic cuts.  We will investigate not only the
magnitude of the DPS cross sections, but also the rapidity
correlations for the DPS $W^\pm W^\pm$ final state, both for the GS09
dPDFs and for a simple factorised PDF model. Note that the recent CDF
and D0 measurements are not accurate enough to distinguish these: the
CDF result \cite{Abe:1997xk} in particular shows no sign of
$x$-dependence in the $\sigeff$ measured.

In assessing suitable benchmark processes for DPS at LHC, one should
of course choose a signal channel for which the single scattering
background is suppressed.  We will be interested in $W^\pm W^\pm$
final states in which both $W$ bosons decay leptonically, $W \to l
\nu$ with $l=e,\mu$. This results in same-sign dilepton (SSL) signals
which we will investigate in detail.  $\Wpm\Wpm$ production has the
advantage that same-sign single scattering is forbidden at the same
order in the SM, i.e. there is no $ij \to W^+W^+,W^-W^-$ contribution
({\em cf.} $q \bar q \to W^+W^-,ZZ$).  The lowest order `background'
process is $\Wpm\Wpm jj$, which is of order $\order(\aw^4)$ or
$\order(\as^2\aw^2)$.  As we shall see, the presence of extra jets
serves as an efficient tag to veto this background.  The possibility
of using lepton pseudorapidity distribution to enhance the DPS signal
was considered in \cite{Treleani:2008talk,Bartalini:2010su}.  For
these reasons, it was believed that the same-sign dilepton channel
provides a `clean environment' for studying DPS processes.  However
there are other important backgrounds, for example $\Wpm \Zgam$ and
$\bbbar$ production, both of which can produce a pair of same-sign
leptons and missing $E_T$.  The former leads to a SSL signal when the
`wrong' sign lepton from $Z$ decay falls outside the detector
acceptance, whereas for the latter, SSL events result when a neutral
$B^0$ meson undergoes $\bmix$ mixing, followed by leptonic decays.
These backgrounds have not been studied in detail in the context of
di-boson production (a brief discussion of the $\Wpm\Zgam$ background
can be found in \cite{Novoselov:2008talk}).  We will go beyond
comparing the DPS $\Wpm\Wpm$ signal and SPS $\Wpm\Wpm jj$ background
\cite{Kulesza:1999zh,Maina:2009vx,Maina:2009sj} to explore the impact
of a fairly standard choice of lepton cuts on SSL events from both the
signal and backgrounds.

This paper is organised as follows. In the following section we
discuss in detail the calculation of the signal $\Wpm\Wpm\to
l^{\pm}l^{\pm} \nu\nu$ DPS process cross section. We compare total
cross sections and rapidity distributions obtained using GS09 dPDFs
and approximate factorised dPDFs. In Sections \ref{sec:background} and
\ref{sec:numerical} we study a number of important background
contributions to the SSL signal, and investigate to what extent they
can be suppressed through final-state cuts. Our conclusions on the
observability of DPS at the LHC in the $WW$ SSL channel are presented
in Section \ref{sec:discussion}.


\section{Signal processes: leptonic channels of $\Wpm\Wpm$}\label{sec:signal}

The DPS signal consists of two same-sign leptons and missing energy,
coming from the decay of two same-sign $W$ bosons. The leptons are
produced in two simultaneous partonic processes \be ij \to W^\pm +X
\to l^\pm + \nu_l +X.  \ee The predictions for the DPS signal are
calculated according to Eq.~(\ref{eq:DPSmaster}).  Total cross
sections and distributions are obtained for four sets of dPDFs: GS09
from~\cite{Gaunt:2009re}, and factorized dPDFs of the form\footnote{We
  use the notation $t = \ln \mu_F^2$.}
\begin{eqnarray}
D_h^{ab}(x_1,x_2,t) &=&
D_h^a(x_1,t)D_h^b(x_2,t)\theta(1-x_1-x_2)\nonumber \\ &&
\times (1-x_1-x_2)^n \qquad n=0,1,2 \;,
\label{eq:MSTWndef}
\end{eqnarray} hereafter referred to as $\textrm{MSTW}_n$ sets. The sPDFs
$D_h^i(x,t)$ are taken from the MSTW 2008 LO set.  As pointed out in
\cite{Gaunt:2009re}, these factorised sets do not satisfy dDGLAP
evolution or consistent sets of sum rules, although with an
appropriately chosen value of $n$ they can provide a reasonable
approximation to the `exact' GS09 dPDFs.  The factorization scale in
the calculations of the DPS signal is fixed at $\mu_F=M_W$ for all
parton sets.

The partonic cross sections in Eq.~(\ref{eq:DPSmaster}) are calculated
at leading order.  At this level, the transverse momentum ($\pt$)
distribution of the produced $W$ bosons is zero.  When we come to
consider signals and backgrounds with realistic experimental cuts, we
will need to provide realistic descriptions of kinematic
distributions, in particular of transverse quantities like the
leptonic $\pt$ and $\MET$, see Section \ref{sec:numerical}.

All numerical results are evaluated with the following electroweak
input parameters \cite{Amsler:2008zzb}: $M_Z=91.188$ GeV, $M_W=80.398$
GeV, $G_F=0.116637\times 10^{-5}\textrm{~GeV}^{-2}$, $\Gamma_Z=2.50$
GeV and $\Gamma_W=2.14$ GeV.  Other EW couplings are derived using
tree level relations.  The effective branching ratio $\BR(\Wp \to
\mu^+\nu_{\mu})$, using $\Gamma_W$ as the total width, is 0.106.  The
CKM mixing parameters used are $|V_{us}|=0.226$, $|V_{ub}|=0.004$,
$|V_{cd}|=0.230$, $|V_{cb}|=0.041$, and other parameters are obtained
using unitarity constraints.  We take the value of $\sigeff=14.5$~mb,
consistent with the Tevatron measurements.

A comparison of the cross sections for DPS and SPS $WW$ production
processes as a function of collider centre-of-mass (CM) energy $\sqrt
s$ is shown in Fig.~\ref{fig:xsecProc}.  The DPS cross sections are
obtained with the GS09 set of dPDFs. We see immediately that while the
single scattering $q \bar{q} \to W^+W^-,\Wpm Z,ZZ$ cross sections
dominate at all collider energies, the DPS $\Wpm\Wpm$ and SPS
$\Wpm\Wpm jj$ cross sections are comparable in magnitude.  As we shall
demonstrate, with an appropriate set of jet veto cuts the SPS
$\Wpm\Wpm jj$ cross sections can be significantly reduced. We quote
the values of the total cross section for DPS $WW$ production
processes for various CM energies at the LHC in
Table~\ref{tab:xsec_DPSvsCME}.  In Tables~\ref{tab:xsec_DPSvsCME} and
\ref{tab:xsec_DPS} we show also the values of the ratio
$$R\equiv4\frac{\sigma_{\Wp\Wp}\;\sigma_{\Wm\Wm}}{\sigma^2_{\Wp\Wm}}$$
which measures the deviation from the factorisation approach,
Eq.~(\ref{eq:naivefact}).  When $R=1$, factorisation is exact.  We
note that factorisation is broken at the 20\% to 30\% level, and the
approximation improves at the higher collider energies as lower $x$
regions are probed.
 
\begin{table}[t]
  \centering
  \begin{tabular}{|c|ccc|}
    \hline 
    & &  $\sigma_{\rm GS09}$  & \\
    \hline
      & $\sqrt s=$7 TeV& $\sqrt s=$ 10 TeV &$\sqrt s=$ 14 TeV\\
    \hline
    $\Wp\Wm$ &0.107 &0.250 &0.546\\
    $\Wp\Wp$ &0.0640 &0.148 &0.321\\
    $\Wm\Wm$ &0.0317 &0.0793&0.182\\
\hline \hline
 & &  $R$  & \\
    \hline 
    &0.709&0.751&0.784\\
    \hline
  \end{tabular}
  \caption{DPS $WW$ total cross sections (in pb) for $pp$ collisions
    at different CM energies.  The values are obtained by
      first calculating the leptonic cross sections, before dividing
      by the corresponding branching ratios.  All cross sections are
    evaluated using GS09 parton distributions.  The ratio $R$ measures
    deviation from the factorisation approach, as explained in the
    text. }\label{tab:xsec_DPSvsCME}
\end{table}

\begin{table}[t]
  \centering
  \scalebox{0.93}{
  \begin{tabular}{|c|cccc|}
    \hline
     &$\sigma_{\textrm{GS09}}$&$\sigma_{\textrm{MSTW}_0}$&$\sigma_{\textrm{MSTW}_1}$&$\sigma_{\textrm{MSTW}_2}$\\
    \hline
    $\Wp\Wm$ &0.546 &0.496 &0.409 &0.348\\
    $\Wp\Wp$ &0.321 &0.338 &0.269 &0.223\\
    $\Wm\Wm$ &0.182 &0.182 &0.156 &0.136\\
    \hline \hline
    & &  \hspace{2cm}$R$  & & \\
    \hline
    &0.784&1.00&1.00&1.00\\
    \hline
  \end{tabular}
  }  
  \caption{DPS $WW$ total cross sections (in pb) for $pp$ collisions
    at $\sqrt s=14$ TeV evaluated using different dPDFs sets.  The
    values are obtained by first calculating the leptonic cross
    sections, before dividing by the corresponding branching ratios.}
\label{tab:xsec_DPS}
\end{table}

\begin{figure*}[!ht]
  \begin{center}
    \begin{tabular}{cc}
      \subfigure[Positively charged leptons]{
        \scalebox{0.6}{
          \includegraphics{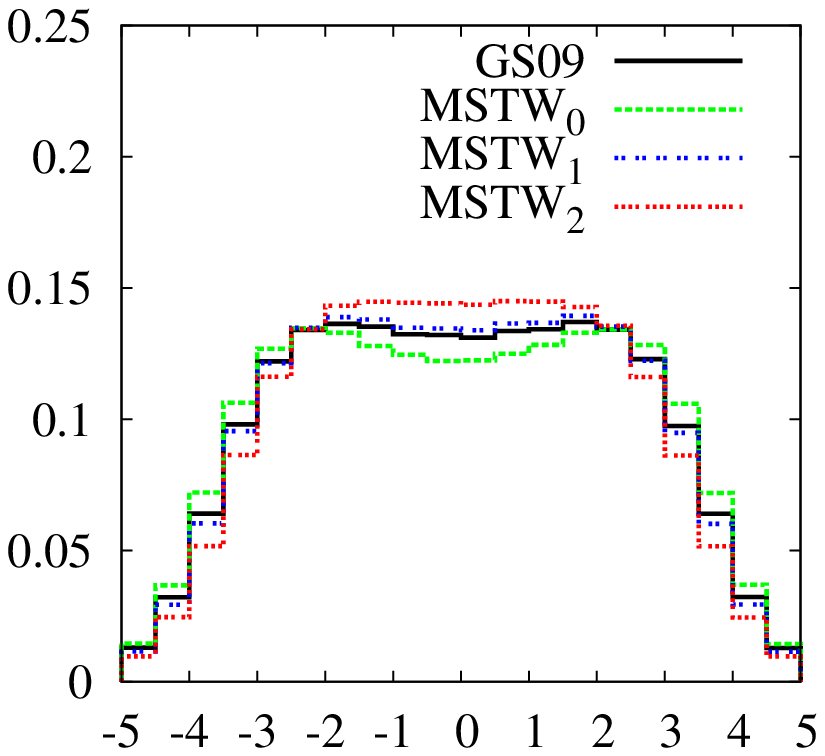}
          \label{fig:yLtot_vs_dpdf+}
        }
        \put(-180,65){\rotatebox{90}{$\frac{1}{\sigma}\frac{\textrm{d}\sigma}{\textrm{d}\eta_l}$}}
        \put(-80,0){$\eta_l$}
        \put(-80,-10){$$}
      }
      &
      \subfigure[Negatively charged leptons]{
        \scalebox{0.6}{
          \includegraphics{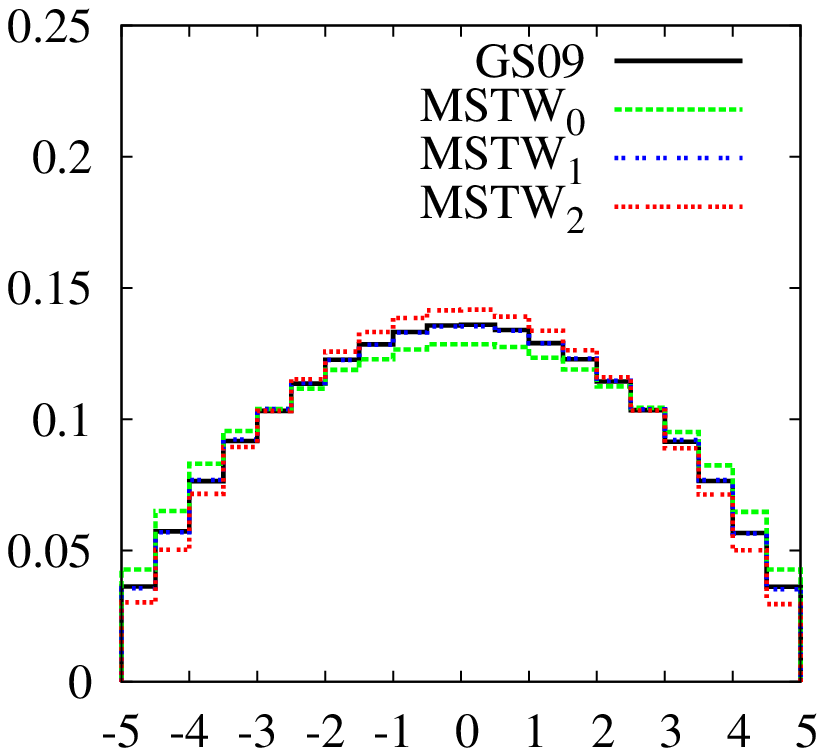}
          \label{fig:yLtot_vs_dpdf-} 
        }
        \put(-180,65){\rotatebox{90}{$\frac{1}{\sigma}\frac{\textrm{d}\sigma}{\textrm{d}\eta_l}$}}
        \put(-80,0){$\eta_l$}
        \put(-80,-10){$$}
      }
      \\
    \end{tabular}
    \caption{Normalised lepton pseudorapidity distributions for $pp$
      collisions at $\sqrt s=14$ TeV evaluated using different dPDFs.
      No cuts are applied.  }
    \label{fig:yLtot_vs_dpdf} 
  \end{center}
\end{figure*}

\begin{figure*}[!ht]
  \begin{center}
    \begin{tabular}{cc}
      \subfigure[Positively charged leptons]{
        \scalebox{0.6}{
          \includegraphics{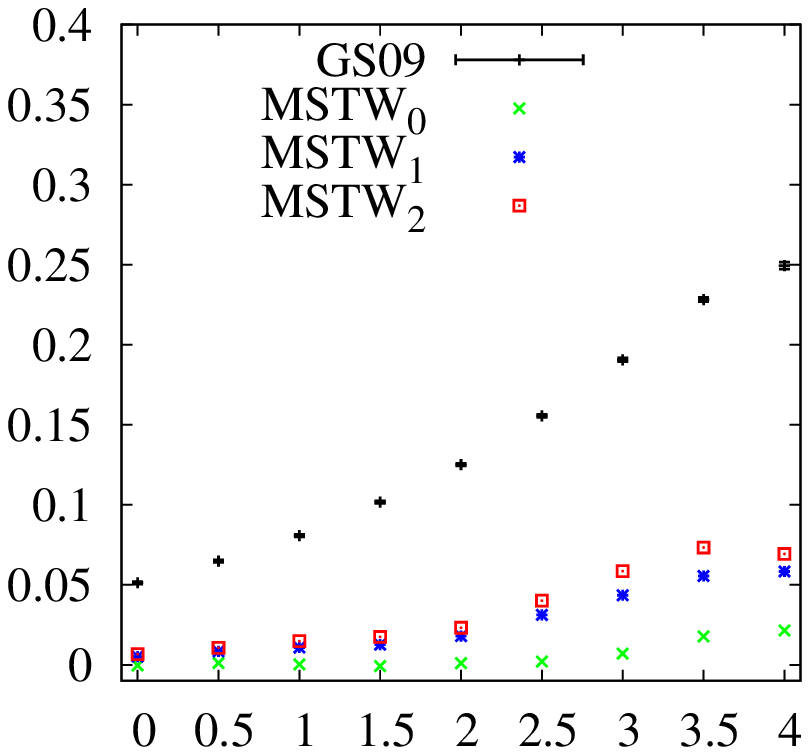}
          \label{fig:AyLtot_vs_dpdf+}
        }
        \put(-175,70){\rotatebox{90}{$a_{\eta_l}$}}
        \put(-80,-5){$\eta_l^{\textrm{\tiny min}}$}
        \put(-80,-15){$$}
      }
      &
      \subfigure[Negatively charged leptons]{
        \scalebox{0.6}{
          \includegraphics{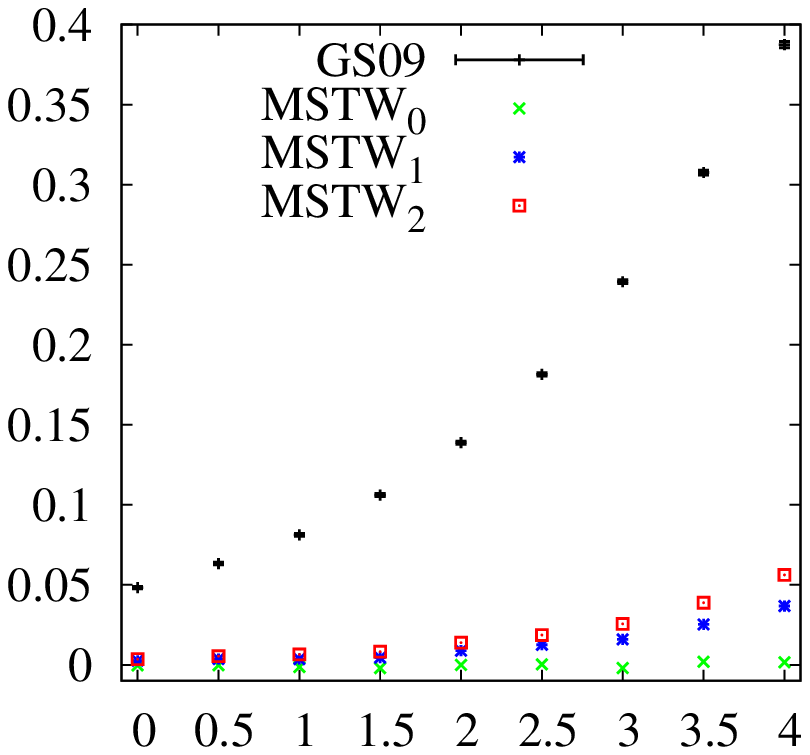}
          \label{fig:AyLtot_vs_dpdf-}
        }
        \put(-175,70){\rotatebox{90}{$a_{\eta_l}$}}
        \put(-80,-5){$\eta_l^{\textrm{\tiny min}}$}
        \put(-80,-15){$$}
      }
      \\
    \end{tabular}
    \caption{Pseudorapidity asymmetry $a_{\eta_l}$ for $pp$ collisions at
      $\sqrt s=14$ TeV evaluated using different dPDFs.  No cuts are
      applied.}
    \label{fig:AyLtot_vs_dpdf}
  \end{center}
\end{figure*}

Since the DPS total cross sections are most relevant for the LHC with
$\sqrt s=14$~TeV, we will focus our analysis on this case. It is
interesting to investigate the effect of the different sets of the
dPDFs, {\it cf.} Table~\ref{tab:xsec_DPS}.  Note that all the
factorised distributions contain a kinematic factor
$\theta(1-x_1-x_2)$, which automatically implies $R \neq 1$ for these
sets, although the effect is numerically very small at LHC (14~TeV)
energies where small $x$ values are probed.  The DPS total cross
sections for same-sign $\Wpm\Wpm$ production are very similar when
using the GS09 and MSTW$_0$ sets whereas the values for the MSTW$_1$
and MSTW$_2$ sets are smaller due to the suppression caused by the
$(1-x_1 - x_2)^n$ factors. The DPS opposite-sign $W^{\pm}W^{\mp}$
cross section is somewhat larger for GS09 than for MSTW$_0$. This is
because of an enhancement of the $D^{q\bar q}$ ($q=u,d)$ GS09 dPDFs
induced by the $g\to q \bar q$ contribution to dDGLAP
evolution~\cite{Gaunt:2009re}.  We also observe that the MSTW$_0$
$\Wp\Wp$ cross section is slightly larger than its GS09 counterpart,
whereas such a difference is not observed in the $\Wm\Wm$ cross
sections. The probable explanation behind this is the following. Both
the $uu$ and $dd$ inputs to GS09 consist of (roughly) factorised forms
minus terms to take account of number effects
($\frac{1}{2}D^{u_v}D^{u_v}$ and $D^{d_v}D^{d_v}$ respectively). In
absolute terms, the subtraction from the $uu$ input is larger than
that from the $dd$ (since $\frac{1}{2}D^{u_v}D^{u_v} \approx
2D^{d_v}D^{d_v}$).  This results in a greater decrease of the GS09
$\Wp\Wp$ cross section relative to that of MSTW$_0$ than occurs for
$\Wm\Wm$ production. In fact, the $\Wm\Wm$ cross section for GS09 ends
up being roughly equal to that of MSTW$_0$, since all GS09 dPDFs are
`fed' by the sPDFs during evolution (either directly or indirectly via
the $gg$ dPDF). The `sPDF feed' contribution to the GS09 dPDFs
compensates for the subtraction from the $dd$ input in the case of the
$\Wm\Wm$ cross section. On the other hand, a similar sPDF feed
contribution cannot counterbalance the greater subtraction from the
$uu$ input in the $\Wp\Wp$ case, causing the GS09 $\Wp\Wp$ cross
section to be smaller than the MSTW$_0$ cross section.  Obviously, the
deviations from the factorization approach measured by the ratio $R$
are small when using the MSTW$_n$ sets. In GS09, there are sizeable
deviations from factorisation, particularly in the valence sector
(i.e. $D^{q_V q_V} \neq D^{q_V}D^{q_V}$), and these are reflected in
smaller values of $R$.

In order to study the potential of using same-sign $WW$ production for
measuring DPS, a full study of the signal, including leptonic decays
and cuts, is necessary.
In Fig.~\ref{fig:yLtot_vs_dpdf+} we present the normalised
pseudorapidity distribution of the $l^+$ leptons, $\frac{1}{\sigma}
\frac{d \sigma}{d \eta_l}$, for different sets of dPDFs.  Although
visible, the differences in the lepton pseudorapidity distributions
for the various dPDFs are small. The effect of the correlations in the
longitudinal momentum fractions in the GS09 set is closely reproduced
by the MSTW$_1$ set, a result which could be expected from the
comparative study of GS09 and MSTW$_n$ distributions
in~\cite{Gaunt:2009re}. Fig.~\ref{fig:yLtot_vs_dpdf-} shows the
corresponding plot for $l^-$ leptons.  Although the shapes of the
pseudorapidity distributions are different (a simple reflection of the
difference between $u$ and $d$ quark PDFs), the qualitative
differences between the various sets for $l^-$ are the same as for
$l^+$ production.

The sensitivity to longitudinal correlations can be maximized in the
following asymmetry \bea
a_{\eta_l}=\frac{\sigma(\eta_{l_1}\times\eta_{l_2}<0)-\sigma(\eta_{l_1}\times\eta_{l_2}>0)}{\sigma(\eta_{l_1}
  \times\eta_{l_2}<0)+\sigma(\eta_{l_1}\times\eta_{l_2}>0)}, \eea
where $\eta_l$ is the lepton pseudorapidity, and
$|\eta_{l_1}|,|\eta_{l_2}|>\eta^{\textrm{min}}_l$. The asymmetry
measures the extent to which the presence of one $W$ produced at high
rapidity affects the probability of finding another $W$ boson with
similarly large rapidity.  A positive $a_{\eta_{l}}$ means that the
leptons prefer to lie in opposite hemisphere.  For higher values of
$\eta_l^{\rm min}$ the effect of the correlations becomes more
pronounced, {\it cf.}  Figs.~\ref{fig:AyLtot_vs_dpdf+}
and~\ref{fig:AyLtot_vs_dpdf-}.  Such behaviour is to be expected as
the correlations are most important for the distributions probed at
high values of $x$ for both partons in the same proton, reached when
the leptons ($W$ bosons) are produced at high rapidities.

In this section we have restricted our attention to basic quantities
(total cross sections and pseudorapidity distributions) to illustrate
the impact of the various dPDF sets, and in particular the effect of
the correlations that are a feature of the GS09 set. As we shall see
below, in practice we have to introduce further cuts on the
final-state particles in order to suppress large backgrounds from SPS
processes. In Section \ref{sec:numerical} we will investigate the
effects on DPS event rates and correlations after imposing these cuts.


\section{Backgrounds}\label{sec:background}

We now turn to discuss background contributions.  Recall that a signal
event consists of two same-sign charged leptons plus missing
transverse energy, where the leptons originate in two separate hard
scatterings from one $pp$ collision.  As already mentioned, $\Wpm\Wpm
jj$ single scattering provides the lowest order, irreducible
background.  Heavy flavour production and gauge boson pair production
can also lead to same-sign dilepton events.  We shall discuss the
relevant properties of these background contributions, and propose
cuts that can reduce them.

\subsection{The $\Wpm \Wpm jj$ single scattering background}
In single scattering, the lowest order in which same-sign $W$'s can be
produced are $\order (\as^2\aw^2)$ and $\order (\aw^4)$. By
considering overall charge conservation, one sees that all the
participating partons must be quarks.  The QCD diagrams involve two
quark propagators and exchange of a $t-$channel gluon, whereas neutral
electroweak gauge boson and Higgs exchanges are also possible for the
EW contributions.  As shown in Fig.~\ref{fig:xsecProc}, the cross
sections for $\Wpm\Wpm jj$ and the DPS $\Wpm\Wpm$ signal are of the
same order of magnitude at LHC energies.

\begin{figure}[!t]
  \begin{center}
    \scalebox{0.7}{ \includegraphics{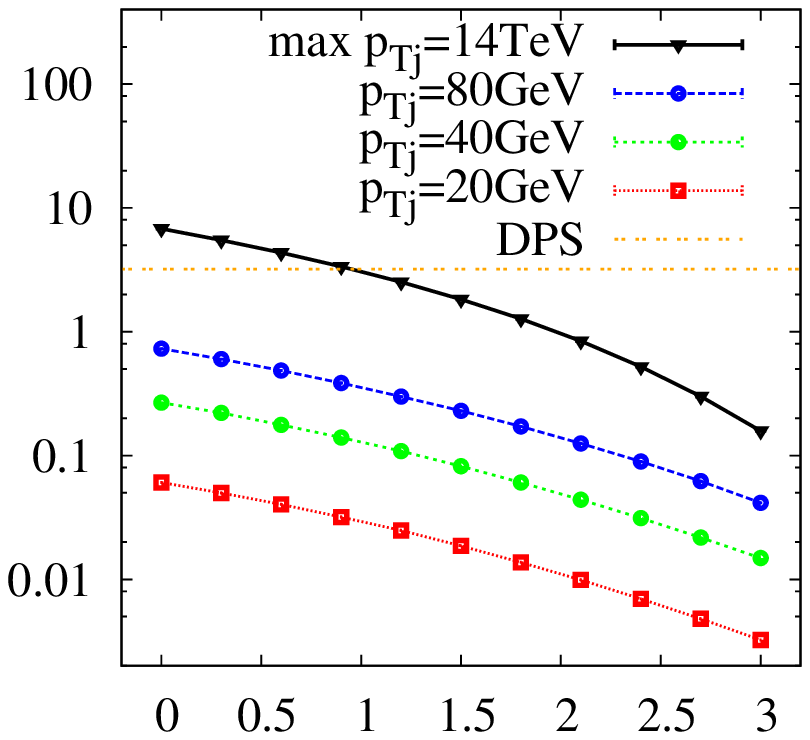} }
    \put(-200,30){\rotatebox{90}{{\footnotesize$\sigma(pp\to2(\Wp\to\mu^+\nu_{\mu})jj)$}
        (fb)}}
    \put(-95,-5){$\eta_j^{\textrm{\tiny min}}$}
    \put(-95,-15){$$}
    \caption{$\sigma(pp\to\Wp\Wp jj)\cdot[\BR (\Wp\to\mu^+\numu)]^2$~(fb) as
      a function of max $p_{Tj}$ and $\eta^{\textrm{\tiny min}}_{j}$.  No
      other cuts are applied.}
    \label{fig:WWjj_PtjmaxYjmin}
  \end{center}
\end{figure}

The presence of jets provide an important handle to suppress this
single scattering background.  The presence of two $W$'s implies that
the jet energies are typically of the order of the weak scale, $M_W$,
and their $\pt$ distribution can have a long tail.  A veto of events
with (central) high $\pt$ jets will therefore be useful in suppressing
this background.  In Fig.~\ref{fig:WWjj_PtjmaxYjmin}, we show the
variation of the cross section of $\sigma(pp\to\Wp\Wp jj)\cdot[\BR
  (\Wp\to\mu^+\numu)]^2$ with maximum transverse momentum (max
$p_{Tj}$) and minimum pseudorapidity ($\eta^{\textrm{min}}_j$) of the
jets allowed in an event.
For example, we see that vetoing events with jet $\pt>20\textrm{ GeV}$
with no pseudorapidity requirement suppresses the cross section to
about two orders of magnitude below the signal (DPS) cross section.
We conclude that this background can be effectively suppressed by a
jet veto, and so will not be discussed further.  However we note in
passing that this process may contribute to SSL backgrounds in beyond
SM scenarios, particularly in event topologies including jets, for
example single slepton production in supersymmetry without R-parity.

\subsection{The heavy flavour background}

The heavy flavour production processes $p p \to Q \bar Q + X$ with
$Q=t,b$ can lead to SSL events. In $\ttbar$ production, the dominant
contribution to same-sign dilepton events is where both a top and the
bottom of the other top decay semi-leptonically, for example \bea t
&\rightarrow& \Wp b \rightarrow l^+\nu b, \nonumber \\ \bar{t}
&\rightarrow& \Wm \bar{b} \rightarrow q\bar q'l^+\nu c.  \eea The
final state is therefore nominally $l^+l^+ + 2\nu + 4$~jets.
Requiring two leptons of the same-sign means that one of the leptons
should come from $b-$quark decay.  This is important because the
$b-$quark, originating from top decay, is energetic and its daughter
lepton will generally not be isolated from the other (hadronic) decay
products.  Hence a tight lepton isolation requirement will be
effective in reducing this background.  The lepton from the $W$
typically acquires a harder $\pt$ spectrum than the signal leptons.
Imposing a maximum lepton $\pt$ cut will therefore also be useful.
Similarly to the $\Wpm\Wpm jj$ process, the presence of energetic jets
again provides an effective suppression mechanism.  Because of these
considerations, we expect that $\ttbar$ production will only
contribute subdominantly to the background, and therefore will also be
neglected in our numerical analysis.

In $\bbbar$ production, the hard process is initiated by two gluons.
To obtain same-sign dileptons, the two resulting $B$ mesons should
decay semi-leptonically.  Further, one of the $B$ mesons must undergo
$\bmix$ mixing before decay, which is possible if the $B$ meson is
neutral.  The relevant processes are thus: \bea gg &\rightarrow&
\bbbar \to B \bar B + ...\; , \nonumber \\ B &\rightarrow& l^+\nu X,
\nonumber \\ \bar{B}^0 &\rightarrow& B^0 \rightarrow l^+\nu \tilde{X},
\eea together with the charge conjugation processes.

Due to its large production cross section at the LHC, this can be a
problematic background (a detailed study of heavy flavour backgrounds
to lepton-pair-plus-missing-transverse-energy final states can be
found, for example, in Ref.~\cite{Sullivan:2006hb}).  Fortunately, the
kinematic properties of this process are very different from those of
the signal, as the two scales involved are very different ($m_b$ {\it
  vs.} $M_W$).  For example, the lepton $\pt$ from $\bbbar$ decay
peaks at very low values and decreases exponentially as $\pt$
increases.  As the $\pt$ of the leptons comes primarily from the $\pt$
of the parent $B$ mesons, at high $\pt$ the lepton will tend to align
with other hadronic decay products making isolation difficult. An
energetic isolated lepton will also tend to be accompanied by a soft
neutrino, leading to low transverse missing energy $\MET$.  As a
result, a combination of lepton isolation, minimum $\pt$ and $\MET$
cuts should effectively suppress this background.  We shall study the
impact of these cuts in the numerical simulation in the following
section.

The $\ccbar$ process also leads to same-sign dilepton events through a
mechanism similar to that of $\bbbar$.  In this case, the lepton $\pt$
spectra peaks at even lower values.  As well as having a strongly
suppressed contribution in our $\pt$ region of interest, the leptons
are again difficult to be isolated from the other decay products, and
low $\MET$ is expected. As a result, the lepton cuts used to suppress
the $\bbbar$ background should effectively suppress $\ccbar$ events as
well.

\subsection{The electroweak gauge boson pair background}

In principle, the production and leptonic decay of heavy weak boson
pairs, $VV \to 4$~leptons with $V=W,Z$ can also provide a sizeable
background.  However these processes do not naturally give rise to
events with same-sign lepton pairs as the only visible particles.  For
example, $\Wpm \Zgam$ and $\Zgam\Zgam$ can lead to 3 or 4 charged
leptons respectively, and can mimic the same-sign di-lepton signal if
the `wrong' sign leptons are not detected.  This happens if they fall
outside the detector acceptance, or if they are not reconstructed.
The relevant processes are then \bea q\bar{q}' &\rightarrow& \Wp \Zgam
\rightarrow l^+\nu l^+(l^-), \nonumber \\ q\bar{q}^{\phantom{'}}
&\rightarrow& \Zgam\Zgam \rightarrow l^+(l^-)l^+(l^-), \eea and their
charge-conjugated processes.  In the above expressions the leptons in
brackets are not identified.  Clearly, a wrong sign lepton veto will
be able to reduce this background when more than 2 leptons are
identified.  The virtual $\gamma^*$ can still contribute significantly
when it decays asymmetrically into a hard and a soft lepton in the
central region.  Following \cite{Chanowitz:1994ap}, this may be
suppressed by looking for isolated charged tracks that form a low
invariant mass with one of the same-sign leptons.  For the $Z$, the
dominant contribution is when the wrong sign lepton lies outside the
central region and is not reconstructed as a lepton.  This in turn
pulls its partner lepton towards the large pseudorapidity region,
potentially providing a shape variable for further discrimination.
Furthermore, the lepton $\pt$ spectrum extends beyond that of the DPS
signal, and therefore a maximum lepton $\pt$ cut will be useful.

\subsection{Other backgrounds}

Another possible source of same-sign lepton pairs comes from
multi-particle interactions, when production of $W$'s of the same sign
from two separate proton collisions occurs during the same bunch
crossing.  In the factorised dPDF approximation, this background and
the signal, where the double scattering occurs in one proton-proton
collision, are expected to exhibit similar kinematic properties.
However the leptons from the multi-particle-interaction background
will in general have tracks pointing back to two different locations
along the beam axis.  In the Appendix, we perform a simple estimation
to show that reasonable longitudinal vertex resolution in the LHC
detectors should allow significant suppression of these events.

Apart from the physics backgrounds discussed, non-physics backgrounds
can also be important.  For example, $W+\;$jets may contribute when a
jet is mis-identified as a lepton, or $ZZ$ production may contribute
when one $Z$ decays invisibly and the charge of a lepton from the
decay of the other $Z$ is mis-identified.  A thorough investigation of
these effects requires a detailed detector simulation.  This is beyond
the scope of the present study, and so we restrict ourselves to
studying the effect of physics backgrounds only.


\section{Numerical Study}\label{sec:numerical}

In the previous two sections we have discussed the DPS signal, and its
characteristic properties, and the most important single scattering
backgrounds. In this section we carry out numerical studies to
investigate the relative size of the various contributions and how to
reduce them.

We first discuss the cuts, referred to as `basic cuts' below, applied
in our numerical analysis.  They are `basic' in the sense that they
are necessary to reduce the background to a manageable level, while
keeping the signal largely intact.  Later we shall discuss refinements
that can improve the signal to background ratio, but at a price of
further reducing the signal.

As discussed in the previous section, the $\Wpm\Wpm jj$ and $\ttbar$
will (the latter partly) be suppressed by a jet veto.  In particular,
rejecting events with jets having $\pt > 20$ GeV will suppress the
$\Wpm\Wpm jj$ background to about two orders of magnitude smaller than
the DPS signal before application of further cuts.  A jet veto is thus
implicitly assumed in the following, and only the $\bbbar$,
$\Wpm\Zgam$ and $\Zgam\Zgam$ backgrounds are simulated.

The basic cuts are as follows:
\begin{enumerate}
\item Both leptons in the like sign lepton pair must have
  pseudorapidity $|\eta|<2.5$.
\item Both leptons are required to be isolated:\\ $E_{\textrm{{\tiny ISO}}}^l\le
  E_{\textrm{{\tiny ISO}}}^{\textrm{{\tiny min}}}=10$ GeV, where $E_{\textrm{{\tiny ISO}}}^l$
  is the hadronic transverse energy in a cone of $R=0.4$ surrounding
  each of the like-sign leptons.
\item The transverse momenta of both leptons, $\ptlep$, must satisfy
  $20 \le \ptlep\le 60$ GeV.
\item An event is rejected whenever a third, opposite-signed, lepton
  is identified.  A lepton is assumed to be identified with 100\%
  efficiency when $\ptlep\ge \pt^{\textrm{\tiny id}}$ and
  $|\eta|<\eta^{\textrm{\tiny id}}$, where $\pt^{\textrm{\tiny id}}=10$ GeV and
  $\eta^{\textrm{\tiny id}}=2.5$.
\item The missing transverse energy $\MET$ of an event must satisfy
  $\MET\ge 20$ GeV.
\item Reject an event if a charged (lepton) track with
  $\pt^{\textrm{\tiny id}}\ge\pt\ge 1$ GeV forms an invariant mass $< 1$ GeV
  with one of the same-sign leptons.
\end{enumerate}
In the above, the muons are treated as invisible particles when
$|\eta|>2.5$.  We define $\MET$ to be the magnitude of the vector
$\pt$ sum of all visible particles.  Note that the above cuts are
designed to be within the capabilities of the LHC general purpose
detectors, ATLAS and CMS. All our cross sections are evaluated at
$\sqrt{s} = 14$~TeV.

We next discuss event simulations.  
The $\bbbar$ events are generated using \herwig \cite{Corcella:2000bw}.
The large total cross section
$\sigma_{\bbbar}\sim 500~\mu\barn$ means that, in practice, a
parton-level cut on the transverse momentum of the bottom quarks,
$\ptb$, is imposed to make the simulation manageable.  This
parton-level cut is chosen to be $\ptb\geq 20$~GeV, which is motivated
by one of the basic cuts, $\ptlep\geq 20$~GeV, discussed above.  The
resulting lepton $\pt$ distribution should remain unchanged, as most
isolated leptons from $B$ meson decays have $\ptlep$ smaller than the
transverse momentum of the parent $b$ quark.  To improve efficiency of
event generation, we further force the $B$ mesons to always decay
semi-leptonically. Whenever one or more neutral $B$ mesons are
produced, exactly one of them undergoes $\bmix$ mixing.  As discussed
in \cite{Dreiner:2000vf}, this neglects the production of leptons in
charm decays.  However these leptons are expected to have a lower
$\pt$ and be less well isolated than the leptons produced in $b$
decays.  The amount of Monte Carlo data generated is equivalent to
about $200$~fb$^{-1}$.  The cross section $\sigma_{\bbbar}(\ptb\geq
20\;\textrm{GeV})$ is then normalised to the corresponding value
obtained in \mcfm~\cite{mcfm} using the central MSTW LO sPDF set.
This value is found to be $\sigma_{\bbbar}(\pt\ge
20\;\textrm{GeV})=5.15\;\mu\barn$, where the renormalisation and
factorisation scales are set at $\mu_R=\mu_F=m_b = 4.75$~GeV
respectively.

Simulations of the other processes are performed at leading-order
parton level only.  Given that the value of $\sigeff$, which controls
the overall magnitude of the DPS signal, is unknown at LHC energies,
this approximation is sufficient for the present purpose.  The parton
distributions and electroweak parameters used are discussed in
Section~\ref{sec:signal}.  The value of $\alpha_S(M_Z)$ is 0.13939,
and we use one loop $\alpha_S$ running throughout.

For $\Wpm \Zgam$ and $\Zgam\Zgam$, we obtain the matrix elements from
\madgraph \cite{Maltoni:2002qb,Stelzer:1994ta}, including both doubly
and singly resonating diagrams.  The phase space integration is
performed using \vegas \cite{Lepage:1977sw}.  All gauge bosons are
decayed leptonically, and off-shell effects are included in the
simulations. The factorisation scale is chosen to be $\mu_F=M_W$ for
all these (leading-order) processes.

The signal processes $\Wpm\Wpm$ are generated in a similar way.  Here
GS09 is used, and again we set $\mu_F=M_W$. The effective cross
section $\sigeff$ is taken to be $14.5$~mb, its value obtained by the
CDF collaboration at the Tevatron.  Since we use cuts on the $\pt$ of
the leptons, it is important to generate realistic $\pt$
distributions.  To account for the non-zero $\pt$ of a $W$ boson, we
introduce a `$\pt$-smearing'.  It is well known that the fixed-order
perturbative calculation fails to describe the $\pt$ distribution of
single gauge bosons produced at small $\pt$ and that the correct
description is provided by the resummed calculations, supplemented by
a parameterization of non-perturbative effects at very small $\pt$.
Resummation takes into account modifications to the $\pt$ distribution
due to multiple soft gluon emission.  We calculate the resummed
distribution to next-to-leading logarithmic (NLL) accuracy with NLO
MSTW pdfs, using the code of Ref.~\cite{Kulesza:1999gm} and the
non-perturbative parameterisation of Ref.~\cite{Kulesza:2003wi}.  The
leptons originating from the two (virtual) $W$ bosons are then boosted
independently in the azimuthal plane according to the $\pt$
distribution obtained in this way.  Even though the dPDFs used in this
study are leading-order quantities, which means that formally it is
more consistent to adopt a leading logarithmic (LL) calculation, we
argue that the described (NLL) procedure is more appropriate for our
studies, since it gives a more realistic description of the $W$ $\pt$
distribution.  However it should be kept in mind that different `$\pt$
smearing' prescriptions will lead to slight changes in kinematic
distributions.

\begin{table}[t]
  \centering
  \begin{tabular}{|c|cc|}
    \hline
    & $\sigma_{\mu^+\mu^+}$ (fb) & $\sigma_{\mu^-\mu^-}$ (fb) \\
    \hline
    $\Wpm\Wpm$(DPS) &0.82 &0.46 \\
    \hline
    $\Wpm \Zgam$ & 5.1 &3.6 \\
    $\Zgam \Zgam$ & 0.84 &0.67\\
    $\bbbar\; (\ptb\geq 20 \textrm{~GeV})$ & 0.43& 0.43\\
    \hline
  \end{tabular}
  \caption[]{Cross sections (in fb) of the processes simulated after
    the basic cuts are applied, including branching ratios
    corresponding to same-sign dimuon production.  The cross sections
    for $\mu^+\mu^+$ and $\mu^-\mu^-$ production are displayed in the
    second and third columns respectively.  The signal cross sections
    are obtained assuming $\sigeff=14.5$~mb.}\label{tab:xsec_bcut}
\end{table}

\begin{figure*}[!ht]
  \begin{center}
      \begin{tabular}{cc}
        \subfigure[$\eta_l$]{ \scalebox{0.65}{
            \includegraphics{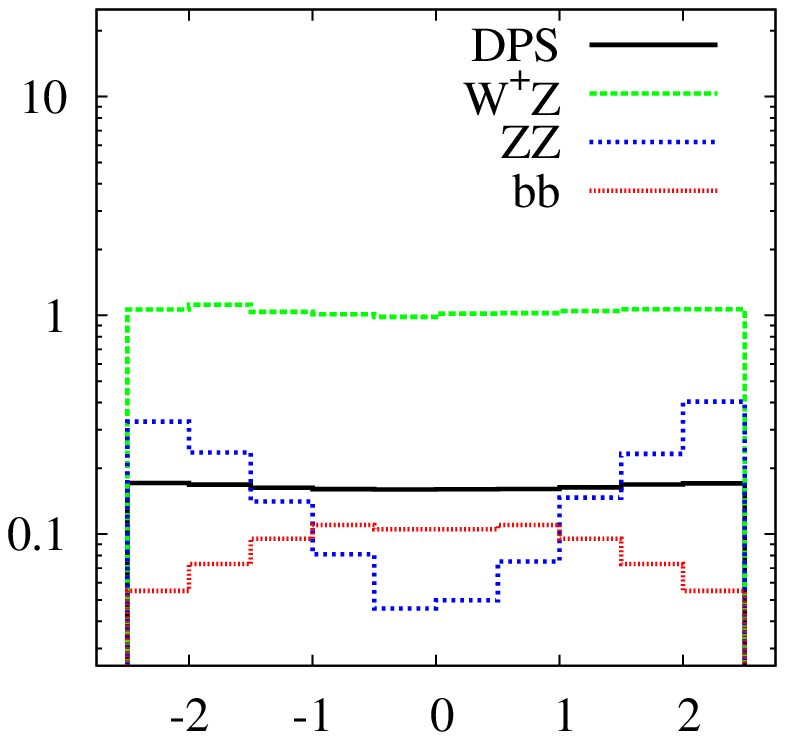} }
          \put(-190,60){\rotatebox{90}{$\frac{\textrm{d}\sigma}{\textrm{d}\eta_l}$(fb/$\eta_l$)}}
        }
        &
        \subfigure[$\Delta\phi_{ll}$]{
          \scalebox{0.65}{
            \includegraphics{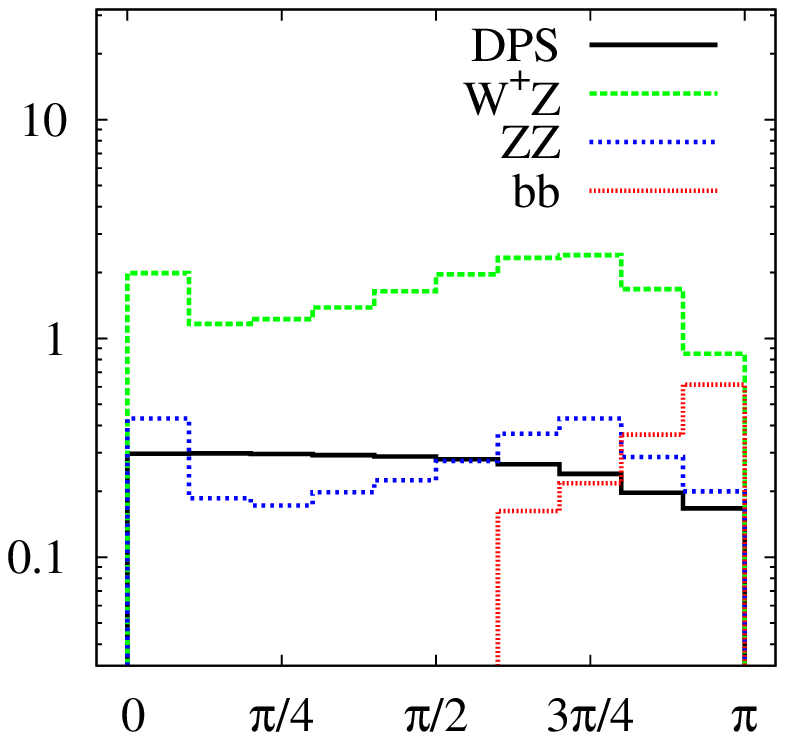}
          }
          \put(-190,60){\rotatebox{90}{$\frac{\textrm{d}\sigma}{\textrm{d}\Delta\phi_{ll}}$(fb/$\Delta\phi_{ll}$)}}
        }
        \\
        \subfigure[$\MET$ (GeV)]{
          \scalebox{0.65}{
            \includegraphics{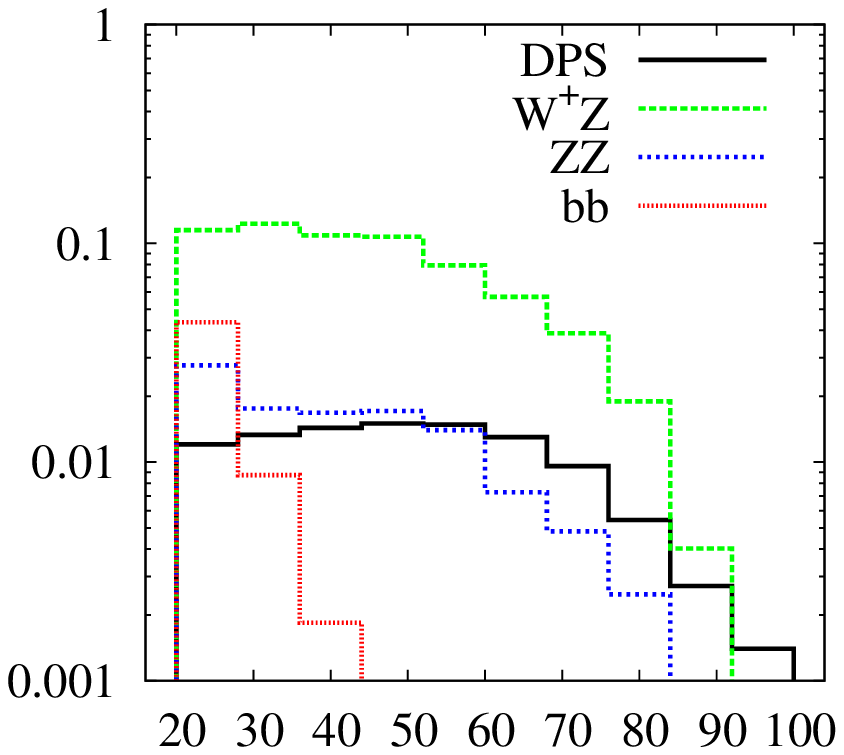}
          }
          \put(-190,60){\rotatebox{90}{$\frac{\textrm{d}\sigma}{\textrm{d}\,\,\MET}$(fb/GeV)}}
        }
        &
        \subfigure[$m_{Tll}$ (GeV)]{
          \scalebox{0.65}{
            \includegraphics{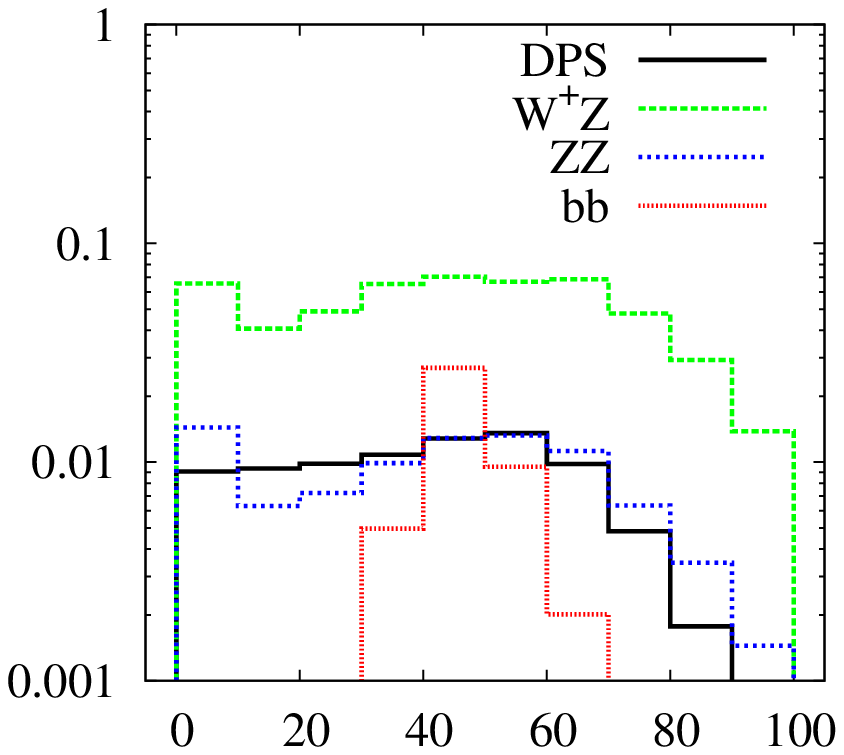}
          }
          \put(-190,60){\rotatebox{90}{$\frac{\textrm{d}\sigma}{\textrm{d}m_T}$(fb/GeV)}}
        }
        \\
        \subfigure[max($p_{Tl1},p_{Tl2}$) (GeV)]{
          \scalebox{0.65}{
            \includegraphics{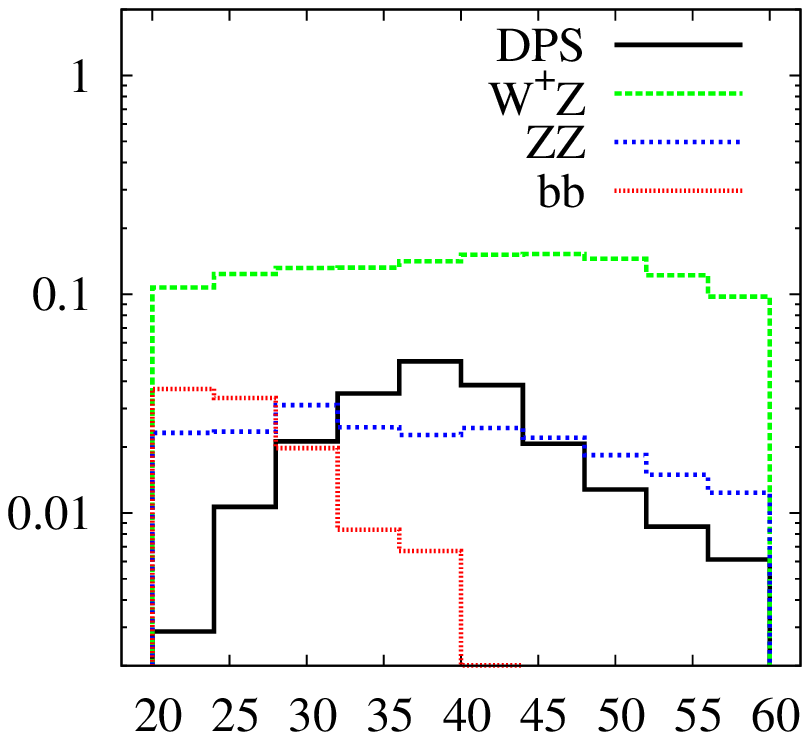}
          }
          \put(-190,60){\rotatebox{90}{$\frac{\textrm{d}\sigma}{\textrm{d}p_T}$(fb/GeV)}}
        }
        &
        \subfigure[min($p_{Tl1},p_{Tl2}$) (GeV)]{
          \scalebox{0.65}{
            \includegraphics{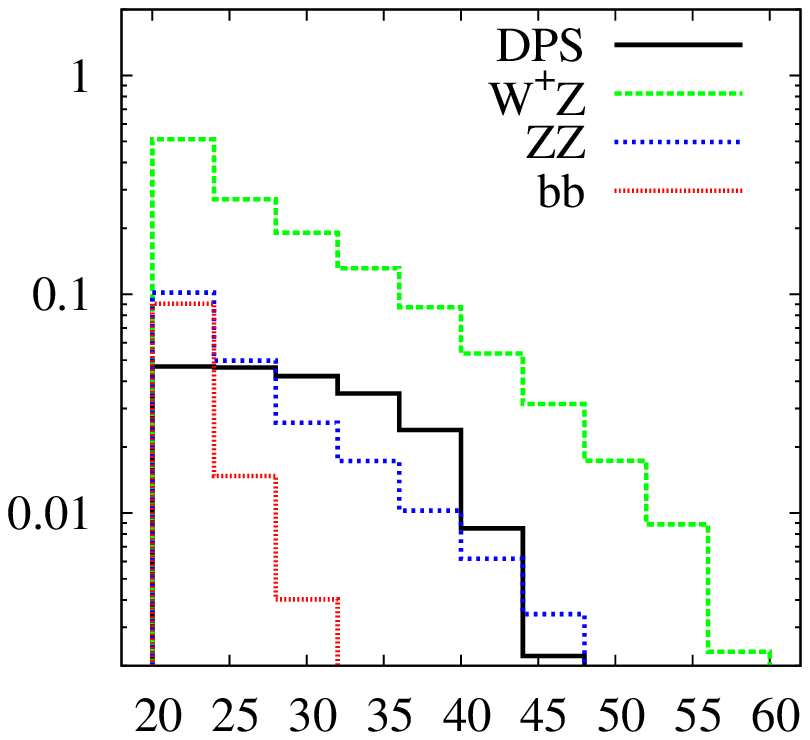}
          }
          \put(-190,60){\rotatebox{90}{$\frac{\textrm{d}\sigma}{\textrm{d}p_T}$(fb/GeV)}}
        }
        \\
      \end{tabular}
      \caption{Selected kinematic distributions for the positively
        charged $(++)$ SSL events.  The DPS signal, $\Wp\Zgam$,
        $\Zgam\Zgam$ and $\bbbar$ backgrounds are described by solid
        black, green dotted, blue dotted and red dashed lines
        respectively.}
      \label{fig:xsecDist+}
  \end{center}
\end{figure*}

The cross sections for the simulated DPS signal and background
processes are displayed in Table~\ref{tab:xsec_bcut}.  The cross
sections correspond to detecting same-sign dimuon events only.
Inclusion of $e^{\pm}e^{\pm}$ and $e^{\pm}\mu^{\pm}$ events can be
estimated by multiplying the results in Table~\ref{tab:xsec_bcut} by
appropriate factors.\footnote{Due to different dPDFs, cuts and lepton
  channels considered, our numerical values for the cross sections are
  different from those presented in Ref.~\cite{Maina:2009sj}.} We see
that although the total $\bbbar$ cross section is huge, the basic cuts
are very effective in suppressing this background.  Instead the
largest background comes from $W\Zgam$ production, whereas the
$\Zgam\Zgam$ background is about a factor of 6 smaller.  In the
latter, we observe an asymmetry of the $\sigma_{\mu^+\mu^+}$ and
$\sigma_{\mu^-\mu^-}$ cross sections.  These two processes would have
the same cross sections if the parton level amplitudes were the same
upon interchanging the momenta of an opposite-sign lepton pair.
However this not the case, as a subset of the Feynman diagrams
contributing to these processes changes sign, leading to different
kinematic (in particular pseudorapidity) distributions.\footnote{It is
  interesting to note that this happens even for diagrams involving
  only $\gamma^*$ and not $Z$.  The parton level amplitudes of these
  diagrams do not change under charge conjugation.  However the
  associated PDF weights change as a result, leading to different
  kinematic distributions at the hadron level.}

\begin{figure*}[!ht]
  \begin{center}
      \begin{tabular}{cc}
        \subfigure[$\eta_l$]{ \scalebox{0.65}{
            \includegraphics{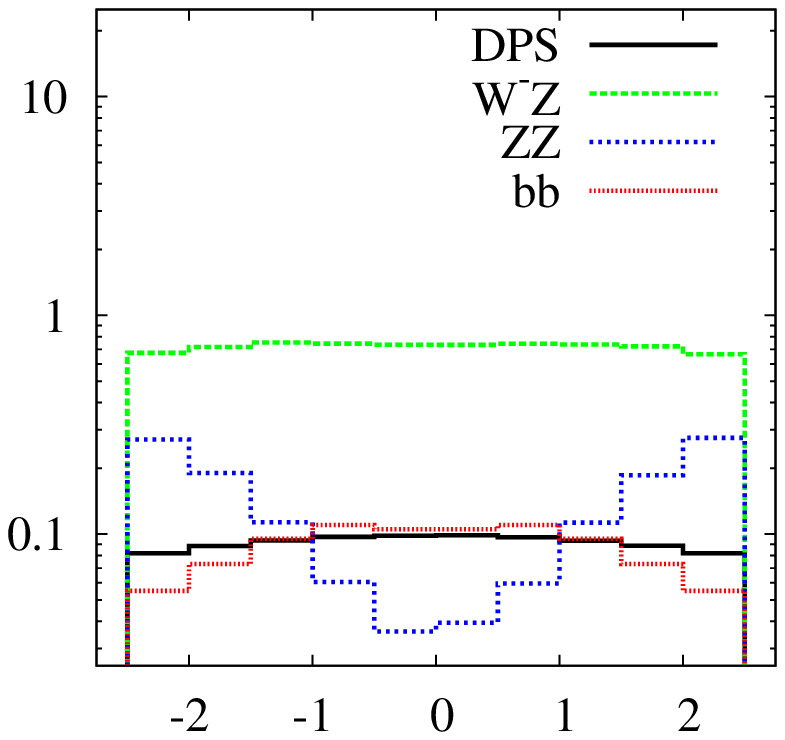} }
          \put(-190,60){\rotatebox{90}{$\frac{\textrm{d}\sigma}{\textrm{d}\eta_l}$(fb/$\eta_l$)}}
        }
        &
        \subfigure[$\Delta\phi_{ll}$]{
          \scalebox{0.65}{
            \includegraphics{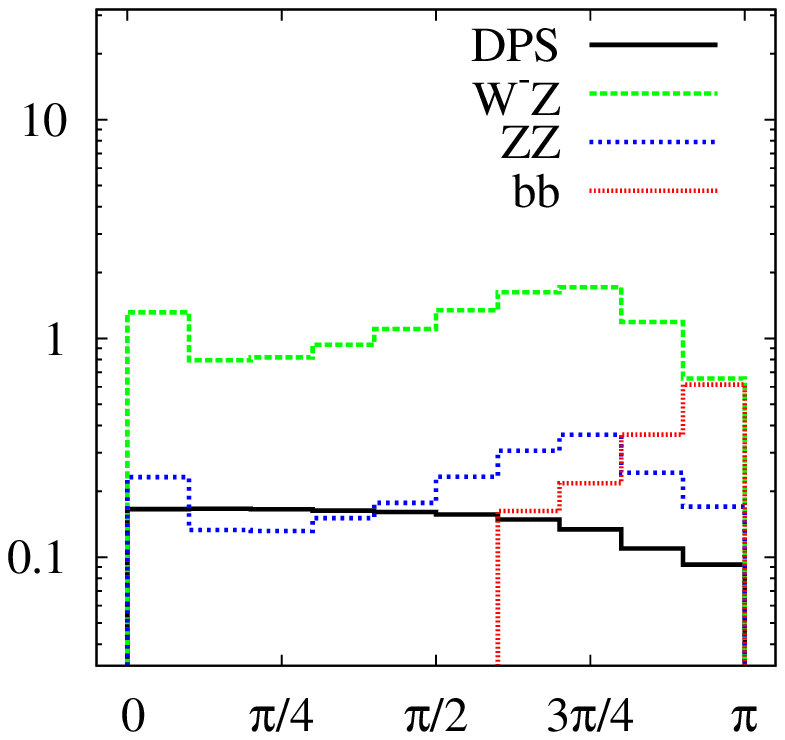}
          }
          \put(-190,60){\rotatebox{90}{$\frac{\textrm{d}\sigma}{\textrm{d}\Delta\phi_{ll}}$(fb/$\Delta\phi_{ll}$)}}
        }
        \\
        \subfigure[$\MET$ (GeV)]{
          \scalebox{0.65}{
            \includegraphics{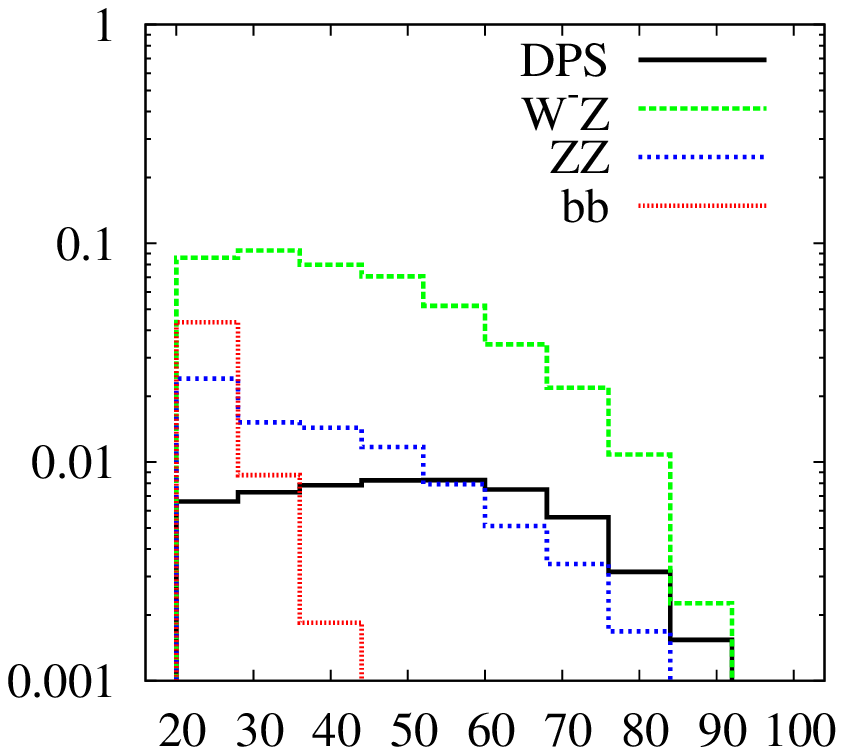}
          }
          \put(-190,60){\rotatebox{90}{$\frac{\textrm{d}\sigma}{\textrm{d}\,\,\MET}$(fb/GeV)}}
        }
        &
        \subfigure[$m_{Tll}$ (GeV)]{
          \scalebox{0.65}{
            \includegraphics{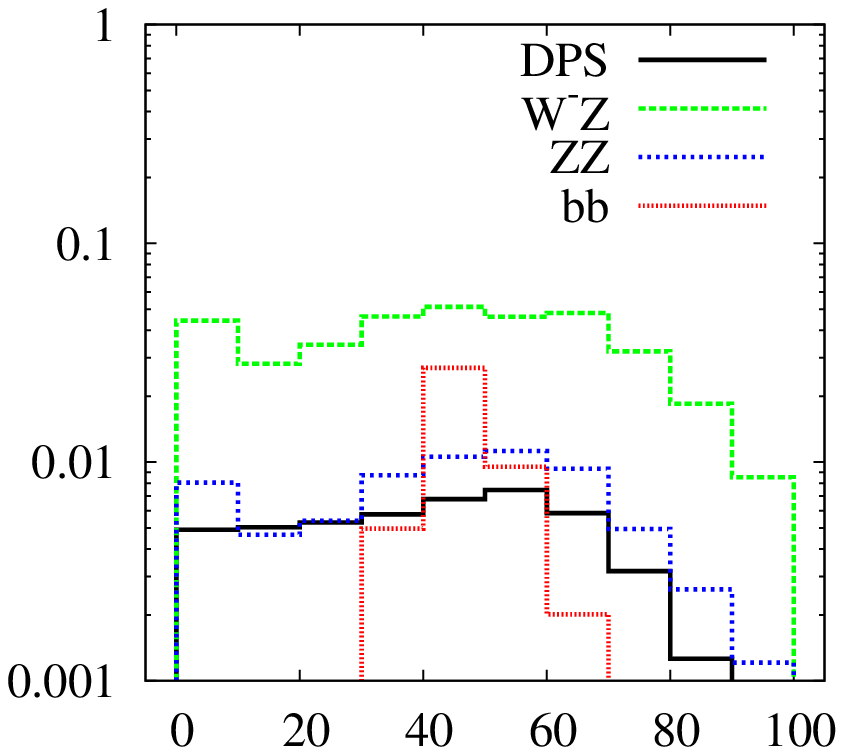}
          }
          \put(-190,60){\rotatebox{90}{$\frac{\textrm{d}\sigma}{\textrm{d}m_T}$(fb/GeV)}}
        }
        \\
        \subfigure[max($p_{Tl1},p_{Tl2}$) (GeV)]{
          \scalebox{0.65}{
            \includegraphics{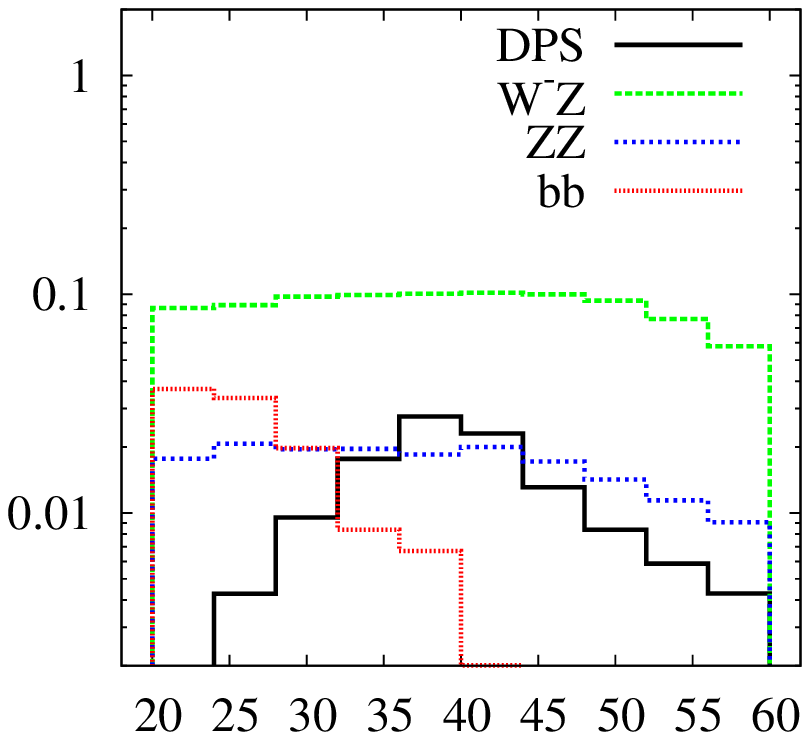}
          }
          \put(-190,60){\rotatebox{90}{$\frac{\textrm{d}\sigma}{\textrm{d}p_T}$(fb/GeV)}}
        }
        &
        \subfigure[min($p_{Tl1},p_{Tl2}$) (GeV)]{
          \scalebox{0.65}{
            \includegraphics{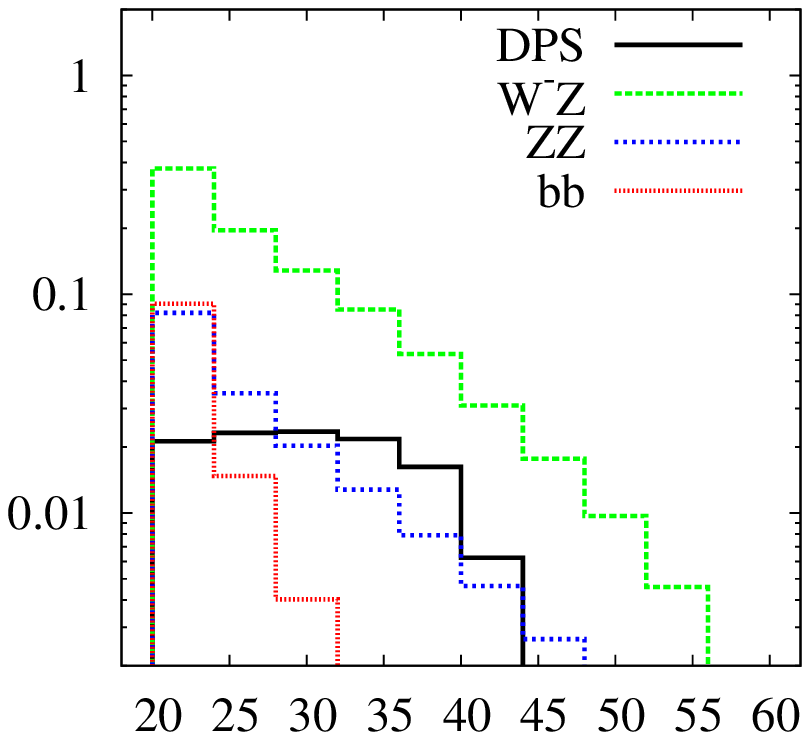}
          }
          \put(-190,60){\rotatebox{90}{$\frac{\textrm{d}\sigma}{\textrm{d}p_T}$(fb/GeV)}}
        }
      \end{tabular}
      \caption{Selected kinematic distributions for the negatively
        charged $(--)$ SSL events.  The DPS signal, $\Wm\Zgam$,
        $\Zgam\Zgam$ and $\bbbar$ backgrounds are described by solid
        black, green dotted, blue dotted and red dashed lines
        respectively.}
      \label{fig:xsecDist-} 
  \end{center}
\end{figure*}

One could consider whether kinematic distributions might be able to
further discriminate the signal from background.  Given that the
signal cross sections are only in the 1~fb region, care is required
not to significantly suppress the signal.

In Fig.~\ref{fig:xsecDist+} and Fig.~\ref{fig:xsecDist-} we show some
representative kinematic distributions for the positively charged
$(++)$ and negatively charged $(--)$ SSL events respectively.  In
these figures, we use only ten bins in each plot to reflect the
anticipated modest number of events expected for the femtobarn level
processes considered, at least in early LHC data taking.

For the signal, the $(++)$ $\eta_l$ distribution is almost flat,
increasing slightly towards large $|\eta_l|$, whereas the signal for
the $(--)$ events shows a small central peak instead.  Both of these
features are characteristic of charged leptons from single Drell-Yan
production.  The distribution of $\Delta\phi_{ll}$, the angular
separation of the SSL pair in the transverse plane, dips near
$\Delta\phi_{ll}=\pi$, as the neutrinos tend to be produced
back-to-back in this limit, thus reducing the missing transverse
momentum $\MET$.

For the $\Wpm\Zgam$ contributions, small peaks appear in the
$\Delta\phi_{ll}=0$ and $m_{Tll}=0$ regions.  Here $m_{Tll}$ is the
transverse mass of the SSL lepton pair.  These two peaks are
correlated, and arise primarily from the $\gamma^*$ contributions.
The $\Zgam\Zgam$ background exhibits similar peaking features for the
same reason.  Furthermore, the $\eta_l$ distribution for this
background peaks strongly towards the high $|\eta|$ region.  This is
because the $Z$ contributes most significantly when one of its
daughter leptons is produced outside the central tracking region,
pulling its partner forward as a result.

Overall, we see that while the $\bbbar$ distributions are distinct
from those of the signal, the dominant $\Wpm\Zgam$ background has
kinematic distributions fairly similar to the signal.  A cut on events
when the lepton pair are approximately back-to-back in the transverse
plane, for example $\Delta\phi_{ll}\lesssim 0.7\pi$, could help, but
further kinematic cuts are unlikely to improve the signal to
background ratio significantly.  Optimisation of the cuts would in any
case require a more detailed detector simulation, and is beyond the
scope of this work.

\begin{figure*}[!ht]
  \begin{center}
      \begin{tabular}{cc}
        \subfigure[+ve charged SSL events]{
          \scalebox{0.65}{
            \includegraphics{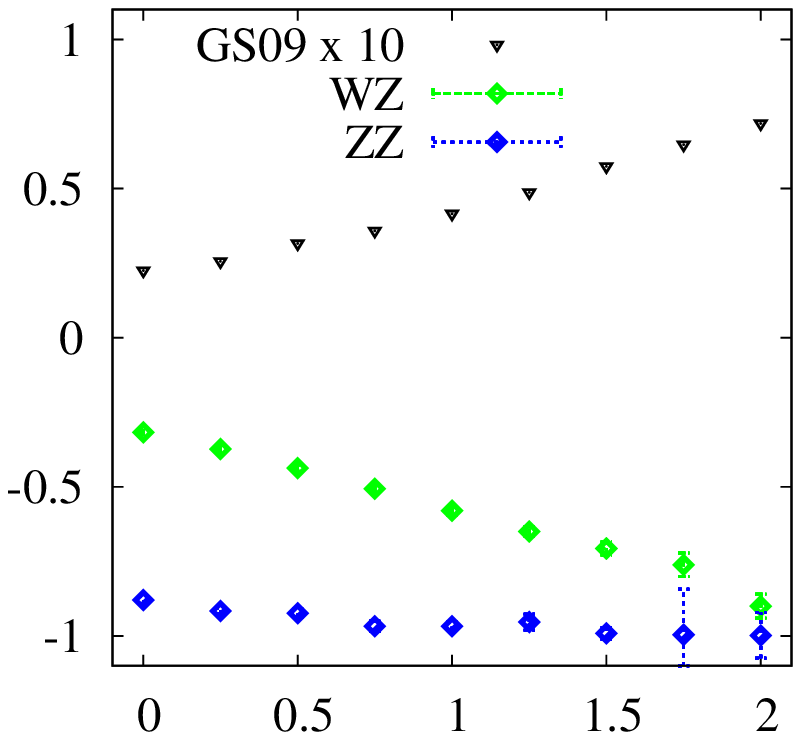}
          }
          \put(-180,80){\rotatebox{90}{$a_{\eta_l}$}}
          \put(-85,-5){$\eta_l^{\textrm{{\tiny min}}}$}
          \put(-85,-15){$$}
        }
        &
        \subfigure[-ve charged SSL events]{
          \scalebox{0.65}{
            \includegraphics{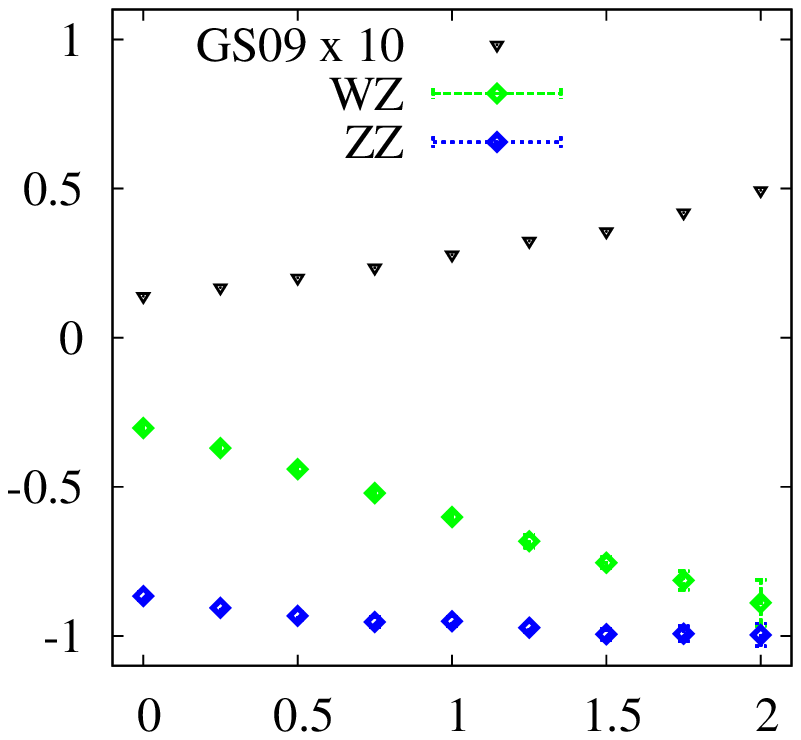}
          }
          \put(-180,80){\rotatebox{90}{$a_{\eta_l}$}}
          \put(-85,-5){$\eta_l^{\textrm{{\tiny min}}}$}
          \put(-85,-15){$$}
        }
      \end{tabular}
      \caption{Pseudorapidity asymmetry $a_{\eta_l}$ as a function of
        $\eta_l^{\textrm{{\tiny min}}}$.  From top to bottom, the points
        correspond to the DPS signal using GS09 and the dominant
        backgrounds $\Wp\Zgam$ and $\Zgam\Zgam$.  For the factorised
        models $\textrm{MSTW}_i$, $a_{\eta_l}$ is practically zero.  For
        clarity, the $a_{\eta_l}$ values for the signal are multiplied
        by a factor of 10.  }
      \label{fig:Ayl_all}
  \end{center}
\end{figure*}

\begin{figure*}[!ht]
  \begin{center}
      \begin{tabular}{cc}
        \subfigure[$\sigma(\Wp \Zgam)$~(fb)]{
          \scalebox{0.65}{
          \includegraphics{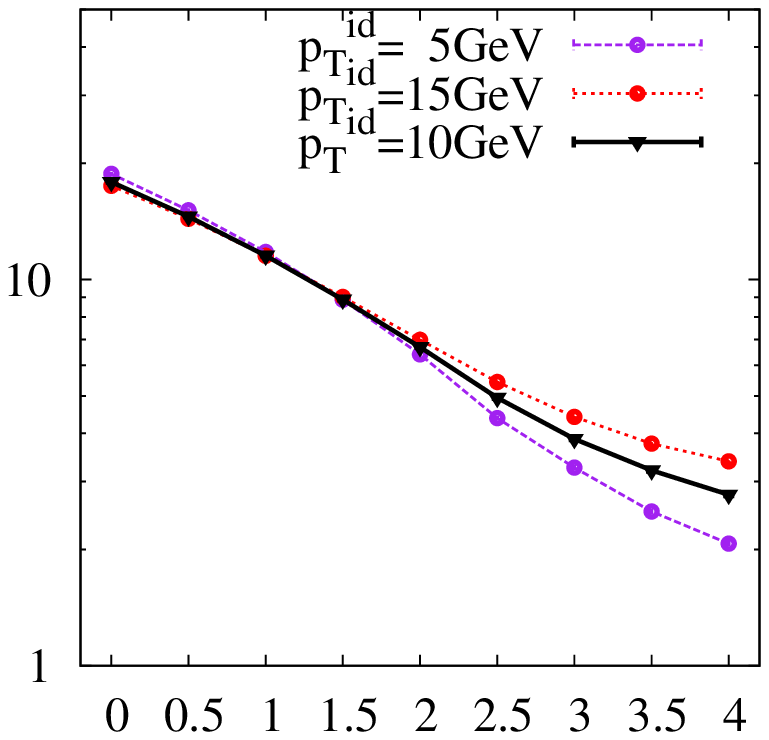}
          }
          \put(-180,75){\rotatebox{90}{$\sigma$(fb)}}
          \put(-85,-5){$\eta^{\textrm{{\tiny id}}}$}
          \put(-85,-15){$$}
        }
        &
        \subfigure[$\sigma(\Wp \Zgam)/\sigma(\Wm \Zgam)$]{
          \scalebox{0.65}{
            \includegraphics{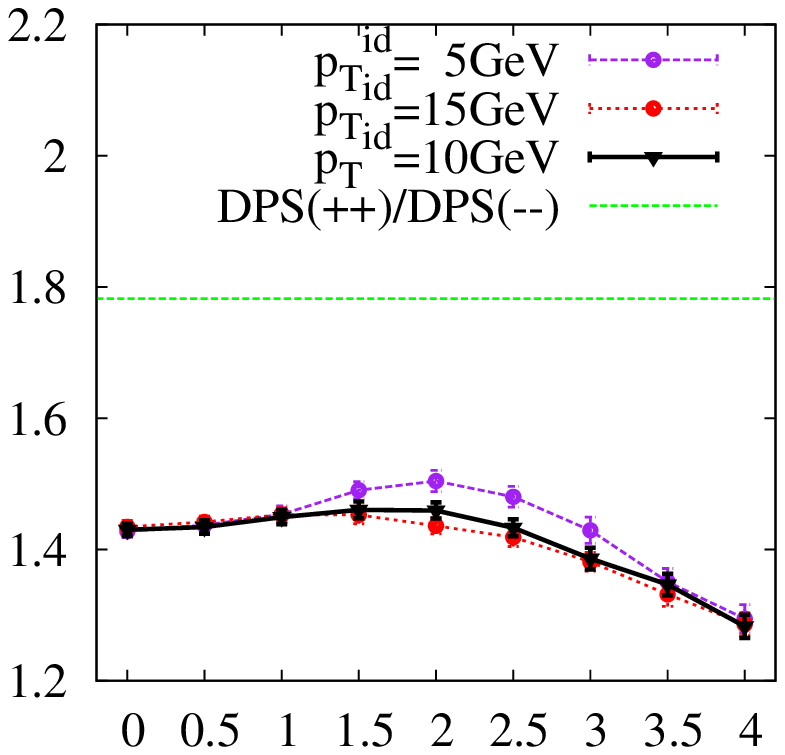}
          }\label{fig:WZRatio}
          \put(-185,75){\rotatebox{90}{$\frac{\sigma(++)}{\sigma(--)}$}}
          \put(-85,-5){$\eta^{\textrm{\tiny id}}$}
          \put(-85,-15){$$}
        }
      \end{tabular}
      \caption{Variation of the $\Wp \Zgam$ cross section and the
        ratio $\sigma(\Wp \Zgam)/\sigma(\Wm \Zgam)$ as a function of
        $\pt^{\textrm{\tiny id}}$ and $\eta^{\textrm{\tiny id}}$.  As discussed in
        the basic cuts in the text, the parameters $\pt^{\textrm{\tiny id}}$
        and $\eta^{\textrm{\tiny id}}$ are the minimum $\pt$ and maximum
        $\eta$ for which a lepton can be identified.  From bottom to
        top, the lines correspond to $\pt^{\textrm{\tiny id}}=$5, 10 and 15
        GeV respectively.  In Fig.~\protect\ref{fig:WZRatio}, the
        ratio for the DPS signal is also shown.}
      \label{fig:WZ_yptVeto}
  \end{center}
\end{figure*}

On the positive side, it might be advantageous to exploit the fact
that the value of $a_{\eta_l}$ is relatively small for the DPS signal.
This means that the probability of the same-sign leptons from the
signal process to lie in opposite ($\eta_1\times\eta_2<0$) and same
($\eta_1\times\eta_2>0 $) hemispheres are approximately equal.
However for the single scattering background, the final states prefer
to have small $\Delta\eta$ in order to minimise the partonic
centre-of-mass energy.  This implies that $a_{\eta_l}$ will generally
be negative.  In Fig.~\ref{fig:Ayl_all}, we show $a_{\eta_l}$ as a
function of $\eta^{\textrm{{\tiny min}}}_l$ after the basic cuts are imposed.  Requiring
the leptons to satisfy $\eta_{l1}\times\eta_{l2}<0$ would reduce the
signal by about a factor of 2, while reducing 2/3 of the $\Wpm\Zgam$
and almost all the $\Zgam\Zgam$ backgrounds.

The ratio of positively charged $(++)$ and negatively charged $(--)$
SSL events may also be used.  Due to the (quasi-)factorised nature of
DPS production, the signal cross ratio
$\sigma^{DPS}_{\Wp\Wp}$/$\sigma^{DPS}_{\Wm\Wm}$ is expected to be
similar to $(\sigma_{\Wp}/\sigma_{\Wm})^2$.  The former ratio can be
inferred from Table~\ref{tab:xsec_DPS} to be about 1.8 at 14~TeV,
whereas the square of $\sigma_{\Wp}$/$\sigma_{\Wm}$ at LO is $\sim
1.36^2=1.85$.  An important point is that this ratio is stable against
the application of cuts, as can be inferred from the results in
Table~\ref{tab:xsec_bcut}.  Because it originates in $gg$ scattering,
the $\bbbar$ background contributes equally to $(++)$ and $(--)$
lepton pair production.  From Fig.~\ref{fig:WZ_yptVeto}, we see that
the $\sigma_{\Wp Z}$/$\sigma_{\Wm Z}$ ratio is roughly 1.4, a
reflection of the $u/d$ parton distribution ratio in the proton.  For
the $\Zgam\Zgam$ process, the corresponding ratio is 1.25 after
including the basic cuts.

Even though these ratios are cut dependent and would presumably change
slightly after properly including higher-order effects, the presence
of DPS events \emph{predicts} that the ratio $(++)$/$(--)$ would
exceed the expectation from single scattering background events only.



\section{Summary}\label{sec:discussion}

We have studied same-sign $W$ pair production as a potential signal of
double parton scattering at 14~TeV at the LHC. We considered both the
DPS signal (in the purely leptonic decay channel) and a number of
backgrounds from the usual single parton scattering.

We first showed that the improved GS09 dPDFs lead to different
production properties compared to naive factorisation models.  In
particular, the use of GS09 dPDFs leads to non-trivial rapidity
correlations of the final-state leptons.  Valence number conservation
implies that it is unlikely to find 2 high-$x$ valence quarks of the
same flavour from a single proton, resulting in a positive lepton
pseudorapidity asymmetry $a_{\eta_l}$ for the DPS processes.
Otherwise, the lepton pseudorapidity distribution using GS09 can be
reasonably well approximated by the $\textrm{MSTW}_1$ quasi-factorized
model.

Our calculations of the DPS signal were undertaken with the assumption
that $\sigeff$ is a constant, and equal to the CDF measured value of
14.5~mb. We remind the reader that although this assumption is
consistent with existing experimental data, it is likely that there
will be a certain degree of process dependence in $\sigeff$
\cite{DelFabbro:2000ds}. It is also possible that different scale
factors might be appropriate for contributions to the dPDF coming from
different terms in the dDGLAP equation \cite{Cattaruzza:2005nu}.
Unless the $\sigeff$ for equal sign WW production at the LHC is very
much smaller than the CDF value, it is likely that the theoretical
aspects just described will be difficult to study in the early stages
of LHC running due to limited statistics. However, the fundamental
characteristic of the DPS signal process -- two (quasi-) independent
scatterings -- means that $\eta_{l1}\times\eta_{l2}$ and the
$(++)$/$(--)$ event ratio could provide additional experimental
handles.


On the other hand, more work needs to be done in order to suppress the
background further.  In addition to the $\Wpm\Wpm jj$ background
previously considered in the literature, we have also considered
di-boson and heavy flavour backgrounds in some detail.  Even though
the `non-reducible' $\Wpm\Wpm jj$ single scattering background can be
effectively suppressed by a central jet veto, the di-boson background
remains significant after applying a fairly basic set of cuts. Using
the canonical $\sigeff=14.5$~mb to calibrate the DPS signal, we find
that the diboson background can be a factor of a few larger than the
signal.

Given the exploratory nature of this study, we have not included
either detector effects or higher-order perturbative corrections.
Given the similarity between the $\Wpm\Zgam$ and signal distributions,
it might not be very beneficial to impose simple additional physics
cuts.  However, the Standard Model processes leading to tri-lepton
events will be studied in detail at the LHC.  An in-depth
understanding of this process might allow accurate extrapolation into
our region of interest, where one of the leptons is not detected.
This would provide an important calibration of the diboson background.
Furthermore, `detector cuts' could also be made.  For example, tighter
lepton isolation and wrong-sign lepton vetos could be used.  Extending
the pseudorapidity region where a (wrong sign) lepton can be
identified could also be useful.  Displaced charged lepton vertices
might also be helpful in further reducing the $\bbbar$ background.

In summary, our preliminary `first look' analysis has shown that a
small excess of SSL events from double parton scattering could be
observed at the LHC, and we are optimistic that further improvements
can be made to enhance the signal.

\begin{acknowledgement}

{\bf Acknowledgements} We are grateful to Richard Batley for helpful
discussions.  A.K.  would like to thank the HEP Group at the Cavendish
Laboratory for warm hospitality and A. M\"uck for useful discussions.
J.G. acknowledges financial support from the UK Science and Technology
Facilities Council.

\end{acknowledgement}

\appendix
\section*{Appendix: multiple particle interaction}
Given a luminosity ($L$), a single scattering cross section ($\sigma$)
and rate of bunch crossing ($B$), the average number of events per
bunch crossing, $\nn$, is given by
\eqa \nn&=&\frac{L\sigma}{B}.
\qea
Using Poisson's statistics, this can be translated into a multiple
particle interaction cross section, $\sigma_N$, where
\eqa \sigma_N &=& e^{-\nn}\frac{\nn^N}{N!}\frac{B}{L} \nonumber \\
&\simeq&\frac{\sigma^N}{N!}\Big(\frac{L}{B}\Big)^{N-1},
\qea
where the last equation holds if $\nn$ $\ll 1$.  An effective
multiple particle interaction parameter, $\sigeffN$, can be
defined analogous to $\sigeff$ for multi-parton interaction:
\eqa
\sigma_N&=&\frac{\sigma^N}{N!(\sigeffN)^{N-1}},\nonumber \\
\sigeffN&\equiv&\Big(\frac{B}{L}\Big).
\qea 

The above approximation ceases to be valid when $\sigma\sim\sigeffN$,
which implies also that $\nn\sim 1$.  For $B=4\cdot 10^{7}s^{-1}$ and
$L=10^{34}\textrm{~cm}^{-2}s^{-1}$, we get $\sigeffN=4 \mb$.  However,
in double particle interactions, the two interaction vertices
typically do not overlap.  Using the RMS bunch length of 7.5~cm
\cite{Zimmermann:2007zza} and intrinsic z-resolution of 115~$\mu$m and
580~$\mu$m for the Pixel detector and SCT at ATLAS \cite{Aad:2009wy}
respectively, the probability that 2 independent scatterings overlap
each other is estimated to be of the order $\mathcal{O}(0.1)$\%.  As a
result this background contribution to double parton scattering,
assuming $\sigeff=14.5\mb$, can be safely neglected.


\begin{thebibliography}{99}

\bibitem{Landshoff:1978fq}
  P.~V.~Landshoff and J.~C.~Polkinghorne,
  Phys.\ Rev.\  D {\bf 18} (1978) 3344.

\bibitem{Takagi:1979wn}
  F.~Takagi,
  Phys.\ Rev.\ Lett.\  {\bf 43} (1979) 1296.

\bibitem{Goebel:1979mi}
  C.~Goebel, F.~Halzen and D.~M.~Scott,
  Phys.\ Rev.\  D {\bf 22} (1980) 2789.

\bibitem{Paver:1982yp}
  N.~Paver and D.~Treleani,
  Nuovo Cim.\  A {\bf 70} (1982) 215.

\bibitem{Humpert:1983pw}
  B.~Humpert,
  Phys.\ Lett.\  B {\bf 131} (1983) 461.

\bibitem{Mekhfi:1983az}
  M.~Mekhfi,
  Phys.\ Rev.\  D {\bf 32} (1985) 2371.

\bibitem{Humpert:1984ay}
  B.~Humpert and R.~Odorico,
  Phys.\ Lett.\  B {\bf 154} (1985) 211.

\bibitem{Ametller:1985tp}
  L.~Ametller, N.~Paver and D.~Treleani,
  Phys.\ Lett.\  B {\bf 169} (1986) 289.

\bibitem{Mekhfi:1985dv}
  M.~Mekhfi,
  Phys.\ Rev.\  D {\bf 32} (1985) 2380.

\bibitem{Halzen:1986ue}
  F.~Halzen, P.~Hoyer and W.~J.~Stirling,
  Phys.\ Lett.\  B {\bf 188} (1987) 375.

\bibitem{Sjostrand:1987su}
  T.~Sjostrand and M.~van Zijl,
  Phys.\ Rev.\  D {\bf 36} (1987) 2019.

\bibitem{Mangano:1988sq}
  M.~L.~Mangano,
  Z.\ Phys.\  C {\bf 42} (1989) 331.

\bibitem{Godbole:1989ti}
  R.~M.~Godbole, S.~Gupta and J.~Lindfors,
  Z.\ Phys.\  C {\bf 47} (1990) 69.

\bibitem{Drees:1996rw}
  M.~Drees and T.~Han,
  Phys.\ Rev.\ Lett.\  {\bf 77} (1996) 4142
  [arXiv:hep-ph/9605430].

\bibitem{Calucci:1997uw}
  G.~Calucci and D.~Treleani,
  Nucl.\ Phys.\ Proc.\ Suppl.\  {\bf 71} (1999) 392
  [arXiv:hep-ph/9711225].

\bibitem{Calucci:1999yz}
  G.~Calucci and D.~Treleani,
  Phys.\ Rev.\  D {\bf 60} (1999) 054023
  [arXiv:hep-ph/9902479].

\bibitem{Akesson:1986iv}
  T.~Akesson {\it et al.}  [Axial Field Spectrometer Collaboration],
  Z.\ Phys.\  C {\bf 34} (1987) 163.

\bibitem{Abe:1997xk}
  F.~Abe {\it et al.}  [CDF Collaboration],
  Phys.\ Rev.\  D {\bf 56} (1997) 3811.

\bibitem{Abazov:2009gc}
  V.~M.~Abazov {\it et al.}  [D0 Collaboration],
  Phys.\ Rev.\  D {\bf 81} (2010) 052012
  [arXiv:0912.5104 [hep-ex]].


\bibitem{DelFabbro:1999tf}
  A.~Del Fabbro and D.~Treleani,
  Phys.\ Rev.\  D {\bf 61} (2000) 077502
  [arXiv:hep-ph/9911358].

\bibitem{DelFabbro:2002pw}
  A.~Del Fabbro and D.~Treleani,
  Phys.\ Rev.\  D {\bf 66} (2002) 074012
  [arXiv:hep-ph/0207311].

\bibitem{Hussein:2006xr}
  M.~Y.~Hussein,
  Nucl.\ Phys.\ Proc.\ Suppl.\  {\bf 174} (2007) 55
  [arXiv:hep-ph/0610207].

\bibitem{Hussein:2007gj}
  M.~Y.~Hussein,
  arXiv:0710.0203 [hep-ph].

\bibitem{Kulesza:1999zh}
  A.~Kulesza and W.~J.~Stirling,
  Phys.\ Lett.\  B {\bf 475} (2000) 168
  [arXiv:hep-ph/9912232].

\bibitem{Maina:2009vx}
  E.~Maina,
  JHEP {\bf 0904} (2009) 098
  [arXiv:0904.2682 [hep-ph]].

\bibitem{Maina:2009sj}
  E.~Maina,
  JHEP {\bf 0909} (2009) 081
  [arXiv:0909.1586 [hep-ph]].

\bibitem{Berger:2009cm}
  E.~L.~Berger, C.~B.~Jackson and G.~Shaughnessy,
  Phys.\ Rev.\  D {\bf 81} (2010) 014014
  [arXiv:0911.5348 [hep-ph]].

\bibitem{Snigirev:2003cq}
  A.~M.~Snigirev,
  Phys.\ Rev.\  D {\bf 68} (2003) 114012
  [arXiv:hep-ph/0304172].

\bibitem{Gaunt:2009re}
  J.~R.~Gaunt and W.~J.~Stirling,
  JHEP {\bf 1003} (2010) 005
  [arXiv:0910.4347 [hep-ph]].

\bibitem{Korotkikh:2004bz}
  V.~L.~Korotkikh and A.~M.~Snigirev,
  Phys.\ Lett.\  B {\bf 594} (2004) 171
  [arXiv:hep-ph/0404155].

\bibitem{Cattaruzza:2005nu}
  E.~Cattaruzza, A.~Del Fabbro and D.~Treleani,
  Phys.\ Rev.\  D {\bf 72} (2005) 034022
  [arXiv:hep-ph/0507052].

\bibitem{Treleani:2008talk}
  D.~Treleani,
  slides presented at MPI at LHC workshop, 27-31 Oct 2008, Perugia, Italy,
  http://agenda.infn.it/getFile.py/access?contribId=8\&\\sessionId=2\&resId=0\&materialId=slides\&confId=599

\bibitem{Bartalini:2010su}
  P.~Bartalini {\it et al.},
  arXiv:1003.4220 [hep-ex].

\bibitem{Novoselov:2008talk}
  A.~Novoselov,
  slides presented at Physics and Computing in ATLAS meeting, 16-19 Sept 2008, IHEP, Protvino,
  http://pcbec3.ihep.su/~miagkov/meet08-2/pr-novoselov-gamma.pdf

\bibitem{Sullivan:2006hb}
  Z.~Sullivan and E.~L.~Berger,
  Phys.\ Rev.\  D {\bf 74} (2006) 033008
  [arXiv:hep-ph/0606271].

\bibitem{Amsler:2008zzb}
  C.~Amsler {\it et al.}  [Particle Data Group],
  Phys.\ Lett.\  B {\bf 667} (2008) 1.

\bibitem{Chanowitz:1994ap}
  M.~S.~Chanowitz and W.~B.~Kilgore,
  Phys.\ Lett.\  B {\bf 347} (1995) 387
  [arXiv:hep-ph/9412275].

\bibitem{Corcella:2000bw}
  G.~Corcella {\it et al.},
  JHEP {\bf 0101} (2001) 010
  [arXiv:hep-ph/0011363].

\bibitem{Dreiner:2000vf}
  H.~K.~Dreiner, P.~Richardson and M.~H.~Seymour,
  Phys.\ Rev.\  D {\bf 63} (2001) 055008
  [arXiv:hep-ph/0007228].

\bibitem{mcfm}
  J.~M.~Campbell and R.~K.~Ellis,
  http://mcfm.fnal.gov/.

\bibitem{Maltoni:2002qb}
  F.~Maltoni and T.~Stelzer,
  JHEP {\bf 0302} (2003) 027
  [arXiv:hep-ph/0208156].

\bibitem{Stelzer:1994ta}
  T.~Stelzer and W.~F.~Long,
  Comput.\ Phys.\ Commun.\  {\bf 81} (1994) 357
  [arXiv:hep-ph/9401258].

\bibitem{Lepage:1977sw}
  G.~P.~Lepage,
  J.\ Comput.\ Phys.\  {\bf 27} (1978) 192.

\bibitem{Kulesza:1999gm}
  A.~Kulesza and W.~J.~Stirling,
  Nucl.\ Phys.\  B {\bf 555} (1999) 279
  [arXiv:hep-ph/9902234];

  A.~Kulesza and W.~J.~Stirling,
  Eur.\ Phys.\ J.\  C {\bf 20} (2001) 349
  [arXiv:hep-ph/0103089].

\bibitem{Kulesza:2003wi}
  A.~Kulesza and W.~J.~Stirling,
  JHEP {\bf 0312} (2003) 056
  [arXiv:hep-ph/0307208].

\bibitem{Zimmermann:2007zza}
  F.~Zimmermann,
  slides presented at LARP Mini-Workshop On Beam-Beam Compensation, 2-4 Jul 2007, Menlo Park, California,
  http://www.slac.stanford.edu/spires/\\find/hep/www?irn=7494106.

\bibitem{Aad:2009wy}
  G.~Aad {\it et al.}  [The ATLAS Collaboration],
  arXiv:0901.0512 [hep-ex].

\bibitem{DelFabbro:2000ds}
  A.~Del Fabbro and D.~Treleani,
  Phys.\ Rev.\  D {\bf 63}, 057901 (2001)
  [arXiv:hep-ph/0005273].

\end{thebibliography}
\end{document}